\documentclass[12pt,a4paper]{article}
\renewcommand{\theequation}{\arabic{section}.\arabic{equation}}

\usepackage{amsmath,amssymb}
\usepackage{graphicx,color}
\usepackage{bbm}
\usepackage{cite}

\def\({\left(}
\def\){\right)}

\newcommand{\nn}{\nonumber}
\newcommand{\Eqn}[1]{&\hspace{-0.5em}#1\hspace{-0.5em}&}
\newcommand{\eqb}{\begin{eqnarray}}
\newcommand{\eqe}{\end{eqnarray}}
\renewcommand{\thefootnote}{\fnsymbol{footnote}}
\newcounter{aff}
\renewcommand{\theaff}{\fnsymbol{aff}}
\newcommand{\affiliation}[1]{
\setcounter{aff}{#1} $\rule{0em}{1.2ex}^\theaff\hspace{-.4em}$}
\def\comma      { \, , }
\def\period     { \, . }

\def\calO   {{\cal O}}
\newcommand{\bbR}{{\mathbb R}}
\newcommand{\bbZ}{{\mathbb Z}}

\def\hatc {\hat{c}}

\begin{document}
\baselineskip=0.7cm
\begin{titlepage}

\hfill\hfill
\begin{minipage}{1.2in}
YITP-11-83  \par\noindent TIT/HEP-614  \par\noindent UTHEP-632  
\end{minipage}

\vspace{0.7cm}
\begin{center}
{\Large \bf
T-functions and multi-gluon scattering amplitudes
}
\lineskip .6em
\vskip1.2cm
{\large Yasuyuki Hatsuda\footnote[1]{\tt hatsuda@yukawa.kyoto-u.ac.jp},
           Katsushi Ito\footnote[2]{\tt ito@th.phys.titech.ac.jp} and
           Yuji Satoh\footnote[3]{\tt ysatoh@het.ph.tsukuba.ac.jp}
            }
\vskip 2em
\affiliation{1} {\normalsize\it Yukawa Institute for Theoretical Physics, Kyoto University\\
Kyoto, 606-8502, Japan} \vskip 1 em
\affiliation{2} {\normalsize\it Department of Physics, Tokyo Institute of Technology\\
Tokyo, 152-8551, Japan} \vskip 1 em
\affiliation{3} {\normalsize\it Institute of Physics, University of Tsukuba\\
Ibaraki, 305-8571, Japan} \vskip 1 em
\vskip 1.5em
\end{center}

\begin{abstract}
\vskip 2ex
\baselineskip=3.5ex

We study gluon scattering amplitudes/Wilson loops in ${\cal N} =4$ super
Yang-Mills theory at strong coupling which correspond to minimal surfaces
with a light-like polygonal boundary in AdS${}_{3}$.
We find a concise expression of the remainder function in terms of
the T-function of the associated thermodynamic Bethe ansatz (TBA) system.
Continuing our previous work on the analytic expansion
around  the CFT/regular-polygonal limit,
we derive a formula of the leading-order expansion 
for the  general $2n$-point remainder function. 
The T-system allows us to encode its momentum dependence in
only one function of the TBA mass parameters,  which is obtained by conformal perturbation theory.  We compute its explicit form in the  single mass cases.
We also find that the rescaled remainder functions at strong
coupling and at two loops are close to each other, and their ratio 
at the leading order approaches a constant near 0.9 for large $n$.

\end{abstract}

%
%
\vspace*{\fill}
\noindent
September  2011

\end{titlepage}
\renewcommand{\thefootnote}{\arabic{footnote}}
\setcounter{footnote}{0}
\setcounter{section}{0}

\baselineskip = 3.5ex

\section{Introduction}
Gluon scattering amplitudes in ${\cal N}=4$ super Yang-Mills theory 
are dual to light-like polygonal Wilson loops \cite{Alday:2007hr,Dr}.
The AdS-CFT correspondence  enables us to 
study these amplitudes/Wilson loops
at strong coupling by calculating the area of the minimal surfaces in AdS
with the same light-like polygonal boundary \cite{Alday:2007hr}.

The logarithm of the ratio of an amplitude to the tree-level 
amplitude is expressed as the sum of the Bern-Dixon-Smirnov (BDS) 
part \cite{Bern:2005iz} and the remainder function part \cite{Alday:2007he}.
The BDS part, which includes the IR divergent
terms,  is determined by the recursive structure of the
amplitudes \cite{Anastasiou:2003kj} and the anomalous dual conformal 
Ward identities \cite{Drummond:2007au}. 
Due to the dual conformal invariance, the remainder function is 
a function of the cross-ratios of external momenta and  has been shown to exist for 
$n(\geq 6)$-point amplitudes  \cite{Drummond:2008aq,Bern:2008ap}.

It is important to determine its exact form 
in order to confirm the AdS-CFT correspondence and study the structure
of the amplitudes.  At weak coupling, it has been 
calculated perturbatively  in some cases  \cite{Anastasiou:2009kna,Brandhuber:2009da,Heslop:2010xa,DelDuca:2009au,DelDuca:2010zg,DelDuca:2010zp,
Goncharov:2010jf,Heslop:2010kq}.
In particular, Heslop and Khoze proposed the two-loop remainder function for
$2n$-point amplitudes for external momenta  in $\bbR^{1,1}$, 
whose kinematical
configuration corresponds to a null polygon with $2n$-cusps 
on the AdS${}_3$ boundary 
\cite{Heslop:2010kq}.
At strong coupling, the remainder function 
can be evaluated by the minimal surfaces in AdS with the help of integrability
\cite{Alday:2009yn,Alday:2009dv,Alday:2010vh,Hatsuda:2010cc,Hatsuda:2010vr,Hatsuda:2011ke,Yang:2010az,Maldacena:2010kp,Bartels:2010ej}.
The cross-ratios of  the cusp coordinates corresponding to 
external momenta therein are  expressed in terms of the Y-functions,
which satisfy the functional relations called the Y-system  \cite{Zamolodchikov:1991et}.  
The remainder function is then written by 
these Y-functions  in addition to the free
energy  associated with the Y-system \cite{Alday:2010vh}.

Under certain asymptotic conditions, the Y-functions also satisfy
the Thermodynamic Bethe Ansatz (TBA) integral
equations, which  describe finite-size effects of
two-dimensional integrable models \cite{Zamolodchikov:1989cf}.
Around the small mass parameter/high-temperature/UV limit,  which
corresponds to regular-polygonal Wilson loops,  one can study the 
free energy as the ground state energy  of the dual channel by using the conformal 
perturbation theory. 
For the minimal surfaces with a  $2n$-gonal boundary  in AdS${}_3$,
the TBA equations are those of the homogeneous 
sine-Gordon (HSG) model \cite{FernandezPousa:1996hi} 
with purely imaginary resonance parameters \cite{Hatsuda:2010cc}.
The relevant CFT  in the UV limit is the generalized parafermion
theory  \cite{Gepner:1987sm} for  ${\rm SU}(n-2)_2/{\rm U}(1)^{n-3}$.  
Similarly for $m$-cusp Wilson loops/amplitudes in AdS${}_4$, 
the TBA equations are those of the HSG model associated with the coset 
${\rm SU}(m-4)_4/{\rm U}(1)^{m-5}$ \cite{Hatsuda:2010cc}.

In order to obtain an  analytic formula for the remainder function around
the CFT point,  which is the main subject of this paper, 
we need to find small (complex) mass expansions 
of  the Y/T-functions.  For this purpose,  
we note that the ratio of the $g$-function (boundary entropy)
\cite{Affleck:1991tk} obeys
the same integral equations as for the T-function \cite{Dorey:2005ak}.
Moreover, the exact $g$-functions for real small masses are 
obtained by the boundary and bulk perturbation of the corresponding
CFT \cite{Dorey:1999cj,Dorey:2005ak}.
Combined with an analytic continuation to complex masses, these give
the analytic expansion  of the T-functions.
The expansion of  the Y-functions is obtained thereof, 
since  the Y-functions are  written generally as ratios  of
the T-functions. Together with the expansion of the free energy, 
the analytic  expansion for the remainder function is derived.

In \cite{Hatsuda:2011ke}, Sakai and the present authors
presented a general formalism to obtain the small-mass expansion of the $2n$-point remainder function for the minimal surfaces in AdS$_{3}$
along the above line of argument.
In particular, the complete leading-order expansion was obtained 
for $n=5$ by examining the single mass cases
 classified in  \cite{Ravanini:1992fi} and  by using the 
exact mass-coupling relations in \cite{Zamolodchikov:1995xk,Fateev:1993av}.
We also derived for $n=4$ an all-order expansion of the integral 
representation of the remainder function obtained in  \cite{Alday:2009yn}.
These expressions were compared with the 2-loop formulas.
After appropriate normalization \cite{Brandhuber:2009da}, 
the two rescaled remainder functions were found to be very close to but 
different from each other.

The purpose of this paper is to study analytically  the remainder function for  the 
minimal surfaces with  a $2n$-gonal boundary in AdS${}_3$ for general $n$.
We  show that the cross-ratios of the cusp coordinates 
appearing in the remainder function are concisely expressed in terms of  
the T-functions of the associated TBA system.  Using this result,
we  derive a formula of the leading-order expansion 
for the general $2n$-point remainder function
at strong coupling. The T-system allows us to encode 
its momentum dependence in only one function of the mass parameters. 
We explicitly compute this function in several simplified cases
where the TBA system contains only one mass scale.
We also compare our  strong coupling results with those at two loops.
As in the case of $n=4, 5$ \cite{Brandhuber:2009da,Hatsuda:2011ke},  
the  rescaled remainder functions \cite{Brandhuber:2009da}
are close to each other, and their ratio at the leading order
decreases to a constant near 0.9 for large $n$.

This paper is organized as follows:
In  section 2, we review some basic properties of the remainder function
and the T- and Y-functions related to the minimal surfaces in AdS${}_3$.
In section 3, we discuss the relation between  the cross-ratios of the
momenta and the T-functions. We then present a formula for
the remainder function expressed in terms of the T-/Y-functions.
In section 4, we  discuss the expansion of the remainder function  
around the CFT point.  In section 5, we compute  the explicit mass parameter 
dependence of the leading expansion  
by using the exact mass-coupling relations in the single mass cases.
In section 6, we compare the remainder function at strong coupling
with the 2-loop formula. We also discuss the large-$n$ limit of 
the expansion of the remainder function.
We conclude with a summary and discussion for future directions in section 7.
Appendix A contains some details about 
the relation between the T-functions and the  cross-ratios 
for even $n$.

\setcounter{equation}{0}
\section{Remainder function for $2n$-point amplitudes}

In this section we introduce the remainder function for the scattering 
amplitudes with external momenta lying in two-dimensional
subspace $\bbR^{1,1}$, which correspond to minimal surfaces in  AdS$_{3}$
\cite{Alday:2009yn,Alday:2010vh, Hatsuda:2010cc,Maldacena:2010kp,Hatsuda:2011ke}.
In this case, the number of gluons should be  even for momentum 
conservation, and thus the boundary  of the minimal surfaces forms a light-like polygon 
with $2n$-cusps on the AdS$_{3}$ boundary.
We label its vertices in the light-cone coordinates as 
$ x_{2k-1}=(x_{k}^+,x_{k-1}^-)$, $ x_{2k}=(x_k^+,x_k^-) $
with identification  $x^{\pm}_{k+n}=x^{\pm}_k$ ($ k=1,\cdots, n$).
The gluon momenta are given by 
\eqb
2 \pi p_{j} = x_{j+1}-x_{j}.
\eqe
In conformal gauge, the equations  for the minimal surfaces reduce to the 
generalized sinh-Gordon equation 
$\partial_z\partial_{\bar{z}}\alpha-e^{2\alpha}+|p(z)|^2e^{-2\alpha}=0$
for a real scalar $\alpha(z,\bar{z})$, where
$(z,\bar{z})$  are worldsheet coordinates. $p(z)$ is 
a polynomial of $z$ of order $n-2$ for a $2n$-sided polygon.
It is convenient to define the coordinate $w$ satisfying $dw=\sqrt{p(z)}dz$ in
order to study the solution.
The area of the minimal surfaces defined by $A=4\int d^2z e^{2\alpha}$ is 
decomposed as 
\begin{equation}
A=A_{\rm sinh}+4\int d^2z\sqrt{p\bar{p}}, \quad
A_{\rm sinh}=4\int d^2z(e^{2\alpha}-\sqrt{p\bar{p}}).
\label{eq:area1}
\end{equation}

In order to calculate the area, it is useful to consider two auxiliary linear 
differential equations for the left and right spinors in AdS${}_3$,  
so that the coordinates of a minimal surface are  constructed as 
products of the two spinors. Introducing the spectral parameter  $\zeta$, 
one can combine these equations into a single linear differential equation.
Its compatibility condition turns out to be the SU(2) Hitchin system, 
which reduces to the above generalized sinh-Gordon equation.
For a polynomial $p(z)$ of order $n-2$, there are $n$ angular regions
called the Stokes sectors, where one can define the large and small solutions.
The small solution in each sector is uniquely defined up to 
a factor. Let $s_j(z,\bar{z};\zeta)$ be the small solution in the $j$-th Stokes
sector, with the  normalized Wronskian
\begin{equation}
\langle s_j, s_{j+1}\rangle\equiv{\rm det}(s_j \; s_{j+1})=1.
\end{equation}
We can extend the index $j$ to take values in integers by analytic
continuation of the solutions with respect to $z$. 
Since the small solutions $s_j$ and
$s_{j+n}$ belong to the same Stokes sector, we have $s_j\propto s_{j+n}$.
We note that from the ratios of the Wronskians 
\begin{equation}\label{Chiijkl}
{\cal X}_{ijkl}(\zeta)={\langle s_i, s_j\rangle \langle s_k, s_l\rangle\over
\langle s_i, s_k\rangle \langle s_j, s_l\rangle},
\end{equation}
one can calculate the cross-ratios of external gluon
momenta as
\begin{equation}\label{CR}
{\cal X}_{ijkl}(1)={x_{ij}^+x^{+}_{kl}\over x^+_{ik}x^{+}_{jl}},\quad
{\cal X}_{ijkl}(i)={x_{ij}^-x^{-}_{kl}\over x^-_{ik}x^{-}_{jl}}.
\end{equation}

We now define the remainder function from the area (\ref{eq:area1}).
The term $A_{\rm sinh}$ in (\ref{eq:area1}) is finite and 
up to a constant turns out to be minus the free energy $A_{\rm free}=-F$ 
of an integrable system, which we will discuss shortly.
The constant term is evaluated in the limit where the zeros of the polynomial $p(z)$
are separated far apart and each zero gives the value for the hexagon.
The second term in (\ref{eq:area1}) diverges since the surface extends to infinity, 
while a finite part is written in terms of the period integrals on the 
Riemann surface $w^2=p(z)$.
Introducing, e.g., the radial cut-off and subtracting the BDS part from the
area, we can obtain the remainder function at strong coupling.
Because of the dual conformal symmetry
\cite{Alday:2007hr,Drummond:2007au,Drummond:2006rz},
it is a function of the cross-ratios of the external momenta. 
To find its functional form is the main problem in this subject.
The formula of the remainder function at strong coupling 
is different for odd $n$ and even $n$ due to 
the monodromy around infinity.

\subsection{Remainder function for odd $n$}

For odd $n$ the remainder function is
\begin{equation}\label{eq:R-odd}
R_{2n}={7\pi\over12}(n-2)+A_{\rm free}+A_{\rm periods}+\Delta A_{\rm BDS}.
\end{equation}
Here $A_{\rm free}$ denotes the free energy part.
$A_{\rm periods}$ is defined by
\begin{equation}
A_{ \rm periods} = i \sum_{r=1}^{(n-3)/2} 
 \bigl( \bar{w}_{r}^{e} w^{m,r} - w_{r}^{e} \bar{w}^{m,r}\bigr),
\end{equation}
where 
$w^{e}_{r}=\oint_{\gamma_r^e}\sqrt{p(z)}dz$ and 
$w^{m,r}=\oint_{\gamma^{m,r}}\sqrt{p(z)}dz$ are
the periods for the electric and magnetic cycles 
with the canonical intersection form
$\gamma_{r}^{e} \wedge \gamma^{m,s} = \delta_{r}^{s}$.
The fourth term is the difference between the BDS formula and 
a part of the area solving the dual conformal Ward identities:
\begin{equation}
\Delta A_{ \rm BDS} 
= \frac{1}{4} \sum_{i,j = 1}^{n}
\log\frac{ c_{i,j}^{+} }{ c_{i,j+1}^{+} } 
\log \frac{ c_{i-1,j}^{-} }{c_{i,j}^{-} },
\label{eq:delabdsodd1}
\end{equation}
where   $c^{\pm}_{i,j}$ are the  sequential cross-ratios 
formed by neighboring distances. To represent  these, we introduce  a notation, 
\begin{equation}
[i_{1},i_{2},i_{3},i_{4},i_{5}, \cdots , i_{2k}]^{\pm} \equiv  
  - \frac{x^{\pm}_{i_{2}i_{3}} x^{\pm}_{i_{4}i_{5}} \cdots  x^{\pm}_{i_{2k}i_{1}}}{
  x^{\pm}_{i_{1}i_{2}} x^{\pm}_{i_{3}i_{4}} \cdots
  x^{\pm}_{i_{2k-1}i_{2k}}},
\label{eq:cr2}
\end{equation}
where $x^{\pm}_{ij}\equiv x_{i}^{\pm} -x_{j}^{\pm}$.
The cross-ratios  are then given by%
\footnote{In the following, we choose the range 
of the indices $i,j$ so that $|j-i| \leq  n$.
} 
\begin{equation}\label{cij}
c^{\pm}_{i,j}=
\left\{
\begin{array}{cl}
[i,i+1,\cdots, j-1,j]^{\pm},& \mbox{$j-i>0$: odd}, \\
\mbox{[}i,i-1,\cdots, j+1,j]^{\pm},& \mbox{$j-i>0$: even} ,
\end{array}
\right. 
\end{equation}
together with $c^{\pm}_{i,j} = c^{\pm}_{j,i}$ and 
$c^{\pm}_{i,i}=c^{\pm}_{i, i+1}=1$.
The path connecting the vertices runs clockwise for odd  $j-i>0$ and
counterclockwise for even $j-i >0 $, respectively. 
 In Fig.~1 (a), we show an example of $c_{1,6}^{\pm}$ for $n=7$.

\begin{figure}[t]
\begin{center}
\resizebox{160mm}{!}{\includegraphics{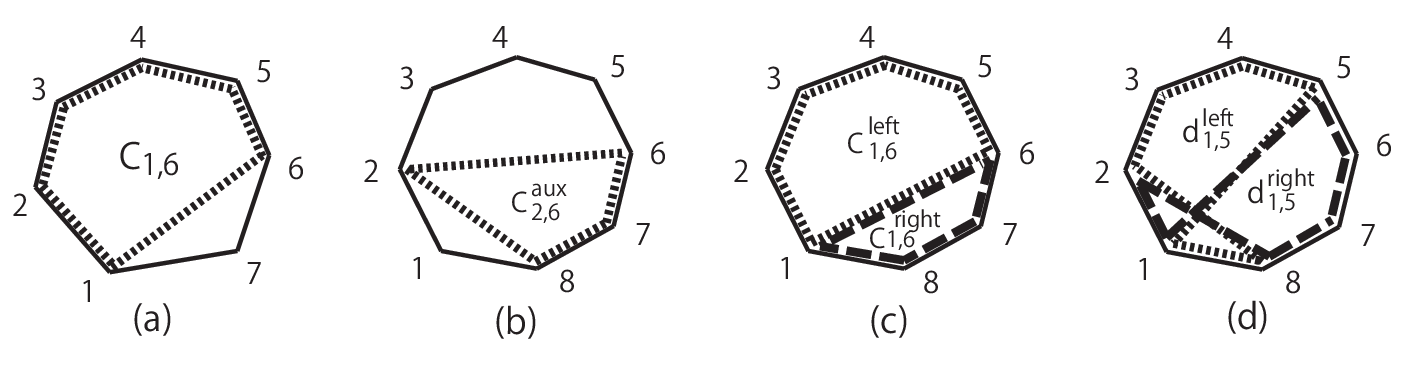}}
\end{center}
\caption{Examples of sequential cross-ratios. $c_{1,6}$ for $n=7$ is shown in (a).
 $c_{2,6}^{aux}$ for $n=8$
is shown in (b). In (c), the dotted and dashed line stand for  $c_{1,6}^{left}$ and
$c_{1,6}^{right}$ for $n=8$, respectively. In (d), 
the dotted and dashed line stand for  $d_{1,5}^{left}$ and
$d_{1,5}^{right}$ for $n=8$, respectively. 
Superscripts $\pm$ are suppressed here for simplicity.}
\label{fig:Cij}
\end{figure}

\subsection{Remainder function for even $n$}

For even $n$ case, the remainder function $R_{2n}$ can be obtained from
the double soft limit of the $2(n+1)$-point amplitudes\cite{Maldacena:2010kp},
which is $x^{\pm}_{n+1}\rightarrow x^{\pm}_1$.
In this limit,  one of the branch point of $\gamma^{m,1}$
is sent to infinity, or equivalently $m_1\rightarrow \infty$
with $m_{2}$ kept finite
in terms of the mass  parameters defined later in (\ref{ms}). 
The remainder function in this case receives contributions from the non-trivial
monodromy around infinity and becomes
\begin{equation}\label{eq:R-even}
R_{2n}={7\pi\over12}(n-2)+A_{\rm free}+A_{\rm periods}
+A_{\rm etxra}+\Delta A_{\rm BDS}.
\end{equation}
Here $A_{\rm free}$ is the free energy again. The period term $A_{\rm periods}$ is
\begin{equation}
A_{ \rm periods} = i \sum_{r=2}^{(n-2)/2} 
\bigl( \bar{w}_{r}^{e} w^{m,r} - w_{r}^{e} \bar{w}^{m,r}\bigr),
\end{equation}
with the same definition of $w_{r}^{e}$, $w^{m,r}$
as  in the  odd $n$ case. 
The extra term $A_{\rm extra}$  is given by
\begin{equation}
A_{ \rm extra}
= -\frac{1}{2} (w_{\rm s} +\bar{w}_{\rm s}) \log \gamma_{1}^{R}
 + \frac{1}{2i} (w_{\rm s} -\bar{w}_{\rm s}) \log \gamma_{1}^{L} ,
\end{equation}
where  $w_{s}$ describes the monodromy of the small solutions 
around infinity and is given by
\begin{equation}\label{ws}
e^{w_{\rm s} +\bar{w}_{\rm s}}
= [1,2,..., n]^{+}, \quad 
 \quad
 e^{(w_{\rm s} -\bar{w}_{\rm s})/i}
= [1,2,..., n]^{-}.
\end{equation}
$\gamma^L_1$, $\gamma^R_1$ are 
the Stokes coefficients of the associated Hitchin equations, 
which are  given by $\langle s_0,s_2\rangle(\zeta)$ at $\zeta=1,i$,  respectively.
Finally, the $\Delta A_{\rm BDS}$ term is given by 
\begin{equation}\label{DelABDSeven1}
\Delta A_{ \rm BDS} = 
\frac{1}{4} \sum_{i,j = 1}^{n+1}
\log\frac{ \hat{c}_{i,j}^{+} }{ \hat{c}_{i,j+1}^{+} } 
\log \frac{ \hat{c}_{i-1,j}^{-} }{\hat{c}_{i,j}^{-} } .
\end{equation}
Here, $\hat{c}^{\pm}_{i,j} = \hat{c}^{\pm}_{j,i}$ $(i,j, = 1, ..., n+1;$ mod $n+1)$ 
are obtained from $c_{i,j}^{\pm}$ for the $2(n+1)$-point amplitudes by the double 
soft limit and take the form,
\eqb
   \hat{c}^{\pm}_{i,j}= \left\{
    \begin{array}{ll}
         c_{i,j}^{aux \, \pm } & (i,j = 2, ..., n) , \\
         c_{1,2k}^{left \, \pm}  & (i=1, \ j=2k),  \\
         d_{1,2k+1}^{right \, \pm}  & (i=1,\  j=2k+1), \\
         c_{1,2k}^{right \, \pm}  & (i=n+1, \, j=2k ),  \\
         d_{1,2k+1}^{left \, \pm}  & (i=n+1, \, j=2k+1) ,
    \end{array}
    \right.
\eqe
with  $\hat{c}^{\pm}_{i,i} = \hat{c}^{\pm}_{i,i+1} =1$.
$c^{aux \, \pm}_{i,j} = c^{aux \, \pm}_{j,i} $ 
are  the cross-ratios for  auxiliary polygons made of
the cusp points  $\{ x_2^{\pm },\cdots, x_{n}^{\pm }\}$: 
\begin{eqnarray}
c^{aux \, \pm}_{i,j}&&=
\left\{
\begin{array}{cl}
[i,i+1,\ldots, j]^{\pm} ,
&
\mbox{$j-i>0$:  odd}, \\
\mbox{[}i,i-1,\ldots,2,n,\ldots,j]^{\pm} , 
&  \mbox{$j-i>0$:  even} ,
\end{array}
\right. 
\end{eqnarray}
with  $c_{i,i}^{aux \, \, \pm} =c_{i,i+1}^{aux \, \pm} =1$. 
$c^{left, right \, \pm}_{1,2k} = c^{left, right \, \pm}_{2k,1}$ and
$d^{left, right \, \pm}_{1,2k+1} = d^{left, right \, \pm}_{2k+1,1}$ 
are the cross-ratios  containing  the vertex $x^{\pm}_{1}$, which are given by
\begin{eqnarray}
c^{left \, \pm}_{1,2k}&= &
[1,2,\ldots,2k]^{\pm} ,
\nn \\
c^{right \, \pm}_{1,2k}&= &
[1,n,\ldots, 2k]^{\pm} , 
\nn \\
d^{left \, \pm}_{1,2k+1}&= &
-[1,n,2,3,\ldots, 2k+1]^{\pm} ,
\\
d^{right \, \pm}_{1,2k+1}&=&
-[1,2,n,n-1,\ldots, 2k+1]^{\pm},  \nn
\end{eqnarray}
and  $c_{1,2}^{left \, \pm} = c_{1,n}^{right \, \pm} =1$.
In Fig.~1 (b)-(d), we show examples of $\hat{c}_{i,j}$ for $n=8$.

\subsection{Y-functions and free energy}

In order to obtain the remainder function as a function of the cross-ratios (\ref{CR}),
we still need to compute $A_{\rm free}$ and $A_{\rm extra}$, 
and to find the relation between $A_{\rm periods}$ 
and the cross-ratios. These are achieved by using  the associated Y- and T-functions. 
In this subsection, we consider $A_{\rm free}$ and $A_{\rm periods}$.
$A_{\rm extra}$ is discussed in the next subsection. 
For a review on the T-/Y-system, see \cite{Kuniba:2010ir} for example. 

For our purpose, we first define the T-functions $T_s$ ($s=0,\cdots, n-2$) by
\begin{equation}\label{defT}
T_{2k+1}(\theta)=\langle s_{-k-1}, s_{k+1}\rangle(\zeta) ,\quad
T_{2k}(\theta)=\langle s_{-k-1}, s_k\rangle(e^{{\pi i \over 2}}\zeta), 
\end{equation}
where  $\zeta=e^{\theta}$  is the spectral parameter. 
{}From the Pl\"ucker relation 
\begin{equation}
\langle s_i, s_j\rangle\langle s_k,s_l\rangle
+\langle s_i,s_l\rangle \langle s_j,s_k\rangle +\langle s_i,s_k\rangle\langle s_l,s_j\rangle=0 ,
\end{equation}
the functions $T_s(\theta)$ are shown to satisfy the T-system of
$A_{n-3}$-type:
\begin{equation}\label{T-system}
T_s\Bigl(\theta +\frac{\pi i}{2} \Bigr)T_s \Bigl(\theta  - \frac{\pi i}{2} \Bigr)
=1+T_{s-1}(\theta) T_{s+1}(\theta), 
\end{equation}
where  $T_{0} =1 $ by definition, 
and one can choose the gauge $T_{n-2} =1$ for odd $n$.
We have also set $T_{-1} =T_{n-1} =0$, which is in accord with (\ref{defT}).
We note that this T-system is invariant under a (residual) gauge transformation
$T_{s} \to e^{\mu_{s} \cosh \theta} T_{s}$ with $\mu_{s}$ being constants
satisfying $\mu_{s+1}+\mu_{s-1} =0$.

We then define the Y-functions $Y_{s}$  ($s=1,\cdots, n-3$) by 
using ${\cal X}_{ijkl}$ in (\ref{Chiijkl}): 
\begin{equation}
Y_{2k}(\theta)=-{\cal X}_{-k,k,-k-1,k+1}(\zeta),\quad 
 Y_{2k+1}(\theta)=-{\cal X}_{-k-1,k,-k-2,k+1}(e^{{\pi i \over 2}}\zeta).
\label{eq:yfunc1}
\end{equation}
These Y-functions satisfy
\begin{equation}\label{YsTs}
Y_s(\theta)=T_{s-1}(\theta)T_{s+1}(\theta),
\end{equation}
and obey the Y-system,
\begin{equation}\label{Y-system}
 Y_s\Bigl(\theta  + \frac{\pi i}{2} \Bigr) Y_s \Bigl(\theta  - \frac{\pi i}{2} \Bigr)
  = \Bigl(1+Y_{s-1} (\theta)\Bigr) \Bigl(1+Y_{s+1}(\theta) \Bigr), 
\end{equation}
Here we have set $Y_{0} = Y_{n-2}=0$, which is in accord with (\ref{eq:yfunc1}).
The  WKB analysis \cite{Gaiotto:2009hg,Alday:2010vh}
shows  that the Y-functions for the minimal surfaces have the asymptotic behavior, 
\begin{eqnarray}\label{Yasympt}
\log Y_s(\theta) & \sim &  -{m_s\over 2\zeta} \quad \  \, (\zeta\rightarrow 0), \nn \\
\log Y_s(\theta) & \sim & -{\bar{m}_s\zeta\over 2} \quad (\zeta\rightarrow \infty).
\end{eqnarray}
Here, we have introduced the ``mass'' parameters $m_{s}$ which are given by 
\begin{equation}\label{ms}
m_{2k}=-2 Z_{2k},\quad m_{2k+1}=2 i  Z_{2k+1}, 
\end{equation}
through the period integrals  $Z_s=-\int_{\gamma_s}\sqrt{p(z)}dz$.
The cycles $\gamma_s$ are related to the electric and magnetic 
cycles $\gamma^e_r$, $\gamma^{m,s}$ by
$ \gamma_{2k}=(-1)^{k+1}(\gamma^e_k-\gamma^e_{k+1}),\ 
\gamma_{2k-1}=(-1)^{k+1} \gamma^{m,k}$.
Their intersection numbers are given by
$\gamma_{2k}\wedge\gamma_{2l-1}=\delta_{k,l}+\delta_{k+1,l}$ and
$\gamma_{2k}\wedge \gamma_{2l}=\gamma_{2k+1}\wedge \gamma_{2l+1}=0$.
Defining the intersection matrix $\theta_{rs} \equiv \gamma_r\wedge \gamma_s$ 
and its inverse $w_{rs} \equiv (\theta^{-1})_{rs}$,  which exits for odd $n$,
the period term $A_{\rm periods}$ takes the form $i w_{rs}Z_{r}\bar{Z}_s$ for odd $n$.
In terms of $m_{s}$, it reads
\begin{align}\label{Aperiodsodd}
A_{\rm periods}=\frac{1}{4} \sum_{k=1}^{{(n-3)}/{2}} \sum_{j=k}^{{(n-3)}/{2}}
(-1)^{j+k+1} (m_{2j}\bar{m}_{2k-1} + \bar{m}_{2j} m_{2k-1}).
\end{align} 
The period term for even $n$ is obtained from $A_{\rm periods}$  for odd 
$n' = n+1$ by the double soft limit: 
\begin{align}\label{Aperiodseven}
A_{\rm periods}=\frac{1}{4} \sum_{k=2}^{{(n-2)}/{2}} \sum_{j=k}^{{(n-2)}/{2}}
(-1)^{j+k+1} (m_{2k-2}\bar{m}_{2j-1} + \bar{m}_{2k-2}m_{2j-1} ).
\end{align}

To derive the integral equations obeyed by the Y-functions,  we introduce
\begin{equation}
\tilde{Y}_s(\theta)=Y_s(\theta+i\varphi_s) ,
\end{equation}
where $\varphi_{s}$ are the phases of the mass parameters,
\begin{equation}
 m_s=|m_s| e^{i\varphi_s} .
\end{equation}
In terms of these $\tilde{Y}_{s }$, the  asymptotic behavior  (\ref{Yasympt}) 
becomes
\begin{equation}
\log \tilde{Y}_s(\theta)\sim -|m_s|\cosh\theta  \quad (|\theta| \to \infty) .
\end{equation}
From the Y-system (\ref{Y-system}), one can then derive the integral equations
\begin{eqnarray}\label{TBA}
\log \tilde{Y}_s(\theta)&&=-|m_s|\cosh\theta+
\int_{-\infty}^{\infty}d\theta' \biggl[
K(\theta-\theta'+i\varphi_s-i\varphi_{s-1})
\log\bigl(1+\tilde{Y}_{s-1}(\theta') \bigr) 
\nonumber\\
&&
+ \, K(\theta-\theta'+i\varphi_s-i\varphi_{s+1})
\log\bigl(1+\tilde{Y}_{s+1}(\theta') \bigr) \biggr],
\end{eqnarray}
where the kernel of the integral is defined by
\begin{equation}
K(\theta)={1\over2\pi}{1\over \cosh\theta}.
\end{equation}
The integral equations  are valid for $| \varphi_{s} - \varphi_{s\pm1}| < \pi/2$,
and are identified \cite{Hatsuda:2010cc} with the TBA equations of 
the homogeneous sine-Gordon  (HSG) model with purely imaginary 
resonance parameters associated with the coset ${\rm SU}(n-2)_{2}/{\rm U}(1)^{n-3}$.

Finally, we obtain $A_{\rm free}$ by using  $\tilde{Y}_{s}$:
\begin{equation}
A_{\rm free}={1\over2\pi}\int_{-\infty}^{\infty}d\theta
\sum_{s=1}^{n-3}|m_s| \cosh\theta \log\bigl(1+\tilde{Y}_{s}(\theta) \bigr).
\end{equation}
In this formalism, the period and free energy terms are given as functions of $m_{s}$. 
These are converted to functions of the cross-ratios through 
the Y-functions,  which are also functions of $m_{s}$. 
Consequently, one obtains  the remainder function for odd $n$ in terms 
of the cross-ratios of external momenta.

\subsection{T-functions and extra term}

To complete the computation of the remainder function for even $n$, we still need 
$A_{\rm extra}$ in terms of the cross-ratios or indirectly of the mass parameters.
It turns out that this reduces to the computation of the T-functions.

In order to obtain the T-functions, we first note that 
the asymptotic behavior of $Y_{s}$ in (\ref{Yasympt}) and the relation 
between the Y- and the T-functions (\ref{YsTs}) lead to the asymptotic behavior,
\begin{eqnarray}\label{Tasympt}
\log T_s(\theta)  &\sim & -{\nu_s\over 2\zeta} \quad \, \, (\zeta\rightarrow 0) , \nn\\
\log T_s(\theta)  &\sim & -{\bar{\nu}_s\zeta\over 2} \quad (\zeta\rightarrow \infty),
\end{eqnarray}
where $\nu_s$ satisfy $m_s=\nu_{s-1}+\nu_{s+1}$.
Since $T_{0} =1$,  one has $\nu_0=0$.
Similarly to the Y-functions, we then introduce 
\begin{equation}
\tilde{T}_s(\theta)=T_s(\theta+i\phi_s),
\end{equation}
where $\phi_{s}$ are the phases of $\nu_{s}$,
\begin{equation}
\nu_s=|\nu_s|e^{i\phi_s}.
\end{equation}
In terms of $\tilde{T}_{s}$, the asymptotic behavior (\ref{Tasympt}) reads
\begin{equation}
\log \tilde{T}_s(\theta)\sim -|\nu_s|\cosh\theta.
\end{equation}
From the T-system (\ref{T-system}), one can then derive the integral equations
\begin{equation}\label{integraleqT}
\log \tilde{T}_s(\theta)=-|\nu_s|\cosh\theta+\int_{-\infty}^{\infty}d\theta'
K(\theta-\theta'+i\phi_s-i\varphi_s)\log\bigl(1+\tilde{Y}_s(\theta') \bigr).
\end{equation}

By solving these equations, one obtains $T_{s}$ as functions of $\nu_{s}$, 
which are in turn expressed by $m_{s}$.
For odd $n$, the gauge $T_{n-2}=1$ implies $\nu_{n-2} =0$. This gives
\begin{eqnarray}
\nu_{2i}=\sum_{k=0}^{i-1}(-1)^k m_{2(i-k)-1} , \quad 
\nu_{2i+1} =\sum_{k=0}^{(n-5)/2-i}(-1)^k m_{2(i+k)+2} .
\end{eqnarray}
For even $n$, the gauge $\nu_{n-2} =0$ is not consistent generally, but   
$\nu_1=0$ is possible instead. With this gauge choice,
\begin{eqnarray}
\nu_{2i} =\sum_{k=0}^{i-1}(-1)^k m_{2(i-k)-1} , \quad 
\nu_{2i+1} =\sum_{k=0}^{i-1}(-1)^k m_{2(i-k)} .
\end{eqnarray}
In particular $\nu_{n-2}$ is given by
\begin{equation}\label{nun2}
\nu_{n-2}=m_{n-3}-m_{n-5}+\cdots +(-1)^{\frac{n}{2}-2}m_1 .
\end{equation}
Since $Y_{n-2}=0$,  we also have 
\eqb\label{Tn2}
\log T_{n-2}(\theta)=
-{1\over2} \bigl(\nu_{n-2}e^{-\theta}+\bar{\nu}_{n-2}e^{\theta} \bigr).
\eqe

Now let us  write down  the extra term $A_{\rm extra}$ for even $n$  
in terms of these T-functions. First, since 
$T_{n-2} = \prod_{k=0}^{n/2 -2} Y^{(-1)^{k}}_{n-3-2k} $, 
the monodromy terms in (\ref{ws}) are given by
\begin{equation}
e^{w_s+\bar{w}_s}= \Bigl(T_{n-2}\Bigl(-{\pi\over2}i \Bigr) \Bigr)^{(-1)^{n/2+1}},\quad
e^{(w_s-\bar{w}_s)/i}= \Bigl(T_{n-2}(0) \Bigr)^{(-1)^{n/2+1}}.
\end{equation}
In addition, the Stokes coefficients $\gamma_{1}^{L,R}$ are given 
through the relations 
$\gamma_1(\zeta)=\langle s_0,s_2\rangle(\zeta) 
= \langle s_{-1}, s_{1}\rangle (e^{\pi i } \zeta)
= T_{1}  (\theta+ \pi i) $.
Putting these together, we find 
\begin{equation}\label{eq:Aextra}
A_{\rm extra}= {(-1)^{{n\over2}}\over2} \biggl[ 
 \log T_{n-2}\Bigl(-{\pi\over2}i \Bigr) \log T_1\Bigl({3\pi\over2} i \Bigr)
-\log T_{n-2}(0)\log T_1(\pi i)
\biggr] .
\end{equation}
By expressing $\nu_{s}$ and $m_{s}$ in terms of  the cross-ratios 
through the Y-functions, one obtains the remainder function for even $n$ 
as a function of momenta.

\subsection{$\bbZ_{2n}$-symmetry and periodicity of Y-/T-functions}

The remainder function is invariant under the cyclic shift  of the
cusp points  $x_{k} \to x_{k+1}$, or in terms of the light-cone coordinates, 
\eqb\label{cyclic}
  x_{j}^{-} \to x_{j+1}^{+} \comma \quad x_{j}^{+} \to x_{j}^{-} .
\eqe
This $\bbZ_{2n}$-symmetry is concisely expressed by the Y-functions as  
\cite{Gaiotto:2010fk},
\eqb\label{Z2n}
  Y_{s} (\theta) \to Y_{s} \Bigl( \theta + \frac{\pi}{2} i\Bigr) .
\eqe
This symmetry strongly constrains the structure of the remainder function 
\cite{Gaiotto:2010fk,Hatsuda:2011ke}.
Moreover, acting with this symmetry twice induces a translation of the light-cone 
coordinates, 
\eqb\label{Zn}
    x_{j}^{\pm} \to x_{j+1}^{\pm} .
\eqe
In the next section, we use this $\bbZ_{n}$-transformation 
for representing cross-ratios by the Y-/T-functions.

Another property used in the later sections is the periodicity of 
the Y-/T-functions.  First, from the Y-system (\ref{Y-system}) with the boundary condition $Y_0=Y_{n-2}=0$,  
one finds  the following half-periodicity of the Y-functions: 
\begin{align}\label{Yperiod}
 Y_s \Bigl(\theta  + \frac{\pi i}{2} n  \Bigr) =Y_{n-2-s}(\theta),
\end{align} 
where $s=0,\dots,n-2$.
We note that this implies the full periodicity $Y_s(\theta+\pi i n)=Y_s(\theta)$.
One can similarly find the periodicity of the T-functions.
For odd $n$, we have the half-periodicity
\begin{equation}
T_s \Bigl(\theta+{\pi i\over 2}n \Bigr)=T_{n-2-s}(\theta),
\label{eq:period-Todd}
\end{equation}
 where $s=0,\dots,n-2$.
For even $n$ one has to take into account the fact that 
the rightmost T-function is not generally equal to unity;  
$T_{n-2}(\theta) \ne 1$. 
Then, the T-system (\ref{T-system}) with the boundary condition
$T_0=1$  and  (\ref{Tn2})
leads to the  quasi-periodicity,
\begin{align}
 T_s\Bigl(\theta  + \frac{\pi i}{2} n  \Bigr) 
 =T_{n-2-s}(\theta)T_{n-2}\Bigl(\theta  + \frac{\pi i}{2} (s-2)  \Bigr),
\label{eq:period-T}
\end{align}
where $s=0,\dots,n-2$ and we have used $T_{n-2}(\theta+ \pi i ) = T_{n-2}^{-1}(\theta)$.
For  the periodicities of the Y- and T-systems, see \cite{IIKNS} for example.

\setcounter{equation}{0}
\section{Cross-ratios and T-functions}

In the previous section, the remainder function was given in terms of 
the Y-/T-functions, the mass parameters specifying their asymptotic 
behavior and the sequential cross-ratios $c_{i,j}^{\pm}, \hat{c}_{i,j}^{\pm}$. 
In this section, we find that $c_{i,j}^{\pm}, \hat{c}_{i,j}^{\pm}$ 
and hence $\Delta A_{\rm BDS}$ 
are concisely expressed by the T-functions. This shows that each term 
in the remainder function is directly represented in the  language of the 
Y-/T-system. Furthermore, it turns out that such a representation enables us 
to derive an analytic expansion of the remainder function around the CFT limit,
beyond numerical analysis or that in the small or large mass limit.

Before going into details, let us summarize our notation.
In terms of  the bracket introduced in (\ref{eq:cr2}), the 4-point
cross-ratios in (\ref{CR}) are given by
\begin{eqnarray}\label{bracketsChi}
  &&  [i,j,k,l]^{\pm} = - \frac{x^{\pm}_{jk} x^{\pm}_{li}}{x^{\pm}_{ij} x^{\pm}_{kl}} 
 = -{\cal X}_{ iljk}(\zeta),
\end{eqnarray}
where $\zeta=1 (i)$ for plus(minus) sign.%
\footnote{
The cross-ratio $\chi_{ijkl} \equiv x_{ij}x_{kl}/x_{ik}x_{jl}$ satisfies
relations such as
$\chi_{ijkl}=\chi_{jilk} = \chi_{klij}$, $\chi_{ikjl} =1/\chi_{ijkl}$, 
$\chi_{lijk}=1-\chi_{ijkl}$, $\chi_{ijkl}/\chi_{ijkm} = \chi_{lkjm}$.
}
From the relation between ${\cal X}_{ijkl}$ and the Y-functions (\ref{eq:yfunc1}), 
we then find that
\begin{eqnarray}\label{bracketsY}
   [k,k+1,-k-2,-k-1]^{+} = Y_{2k+1}^{[-1]},  &&
        [k,k+1,-k-1,-k]^{+} = Y_{2k}^{[0]},  \nonumber \\
     {}[k,k+1,-k-2,-k-1]^{-} = Y_{2k+1}^{[0]}, &&
        [k,k+1,-k-1,-k]^{-} = Y_{2k}^{[1]} , 
\end{eqnarray}
where 
\eqb
   Y^{[k]}_{s} \equiv Y_{s} \Bigl(\frac{\pi i}{2} k \Bigr) .
\eqe
These relations are  understood graphically: 
in the $n$-gons formed by $x^{\pm}_{i}$, the Y-functions at special values of 
$\theta$ are identified with the tetragons which are represented by the brackets in
(\ref{bracketsY}) . 
In Fig.~2, we show an example for $n=7$ and $x_{i}^{+}$. 

\begin{figure}[t]
\begin{center}
\resizebox{100mm}{!}{\includegraphics{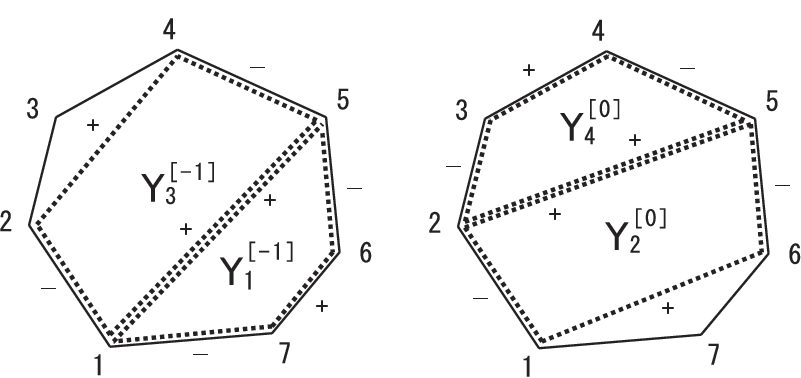}}
\end{center}
\caption{Graphical representation of Y-functions for $n=7$. $Y_s^{[k]}$
are represented by tetragons in the heptagon formed by the cusp coordinates $x^\pm_{i}$
$(i=1, ..., 7)$.  Here, the $i$-th vertex stands for  $x_{i}^{+}$.
 The $+$ sign indicates factors appearing in the numerator of the cross-ratios, whereas
the $-$ sign indicates those in the denominator.
}
\label{fig:yfunction}
\end{figure}

\subsection{Odd $n$ case}

Now, let us discuss the relation between the sequential cross-ratios and
the Y-/T-functions. We begin with the odd $n$ case, where $c^{\pm}_{i,j}$ are given
in (\ref{cij}). We recall that the subscripts $i,j$ labeling the vertex are
defined modulo $n$.

To find the relation of our interest, we first derive  recursion relations 
among $c^{\pm}_{i,j}$. As a simple example, let us consider
$c^{+}_{1,n-2}=c^{+}_{1,-2}=[1,0,-1,-2]^{+}=Y_1^{[-1]}$. By adding two vertices,
one has $c_{2,-3}=[2,1,0,-1,-2,-3]^{+}$. Multiplying 
these two then gives $Y_{3}^{[-1]}$:
\begin{equation}
c^{+}_{1,-2} c^{+}_{2,-3} =[1,2,-3,-2]^{+}=Y_3^{[-1]}.
\end{equation}
This is easily understood graphically as in Fig.~3, 
where $c_{1,-2}^{+}$ and
$Y_{3}^{[-1]}$ are represented as a tetragon whereas $c_{2,-3}$ is as a hexagon.
Continuing similar procedures,  we also have
\begin{equation}
c^{+}_{k,-k-1}c^{+}_{k+1,-k-2}=[k,k+1,-k-2,-k-1]^{+}=Y_{2k+1}^{[-1]},
\label{eq:cy1}
\end{equation}
where  $k=0,\cdots, r-2$  and we have set $ n \equiv  2r+1$.
Another simple example is given by 
$c^{+}_{r-1,-r+1} = Y_{n-3}^{[0]}$. 
Multiplying this with $c^{+}_{r-2,-r+2} $, we have 
$ c^{+}_{r-2,-r+2} c^{+}_{r-1,-r+1} =Y_{n-5}^{[0]}$. 
Similarly, we  find 
\begin{equation}
c^{+}_{k,-k} c^{+}_{k+1,-k-1}= [k,k+1,-k-1,-k]^{+} =Y_{2k}^{[0]},  
\label{eq:cy2}
\end{equation}
where $ k=1,\cdots,r-1 $.

\begin{figure}[t]
\begin{center}
\resizebox{53mm}{!}{\includegraphics{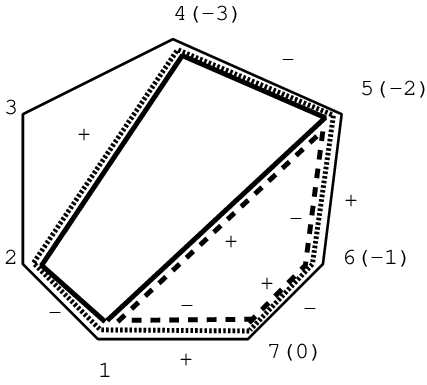}}
\end{center}
\caption{Graphical representation of a recursion relation for $c_{i,j}$. 
The dashed line represents $c_{1,-2}$.
The dotted line stands for $c_{2,-3}$. The bold line represents $Y_3$. }
\label{fig:crodd1}
\end{figure}

Next, we invert the relations (\ref{eq:cy1}) and (\ref{eq:cy2}), to find 
\begin{eqnarray}\label{cr3}
c^{+}_{k+1,-k-2} \Eqn{=} \prod_{l=0}^{k}  \bigl( Y_{2k+1-2l}^{[-1]} \bigr)^{(-1)^{l}}
  =T^{[-1]}_{2k+2}, 
\nonumber\\
c^{+}_{k,-k}  \Eqn{=} \prod_{l=0}^{\frac{n-3}{2}-k}  \bigl(Y_{2k+2l}^{[0]} \bigr)^{(-1)^{l}}
=T_{2k-1}^{[0]}.
\end{eqnarray}
These cover all the non-trivial sequential cross-ratios which contain 
the tetragonal factor 
$[0,1,-2,-1]^{+} = Y_{1}^{[-1]}$ or
$[r-1, r,- r,-r+1]^{+} = Y_{n-3}^{[0]}$. 
To obtain other cross-ratios, we use the $\bbZ_{n}$-transformation 
$x_{j}^{\pm} \to x_{j+1}^{\pm}$ in (\ref{Zn}). 
Since this  is generated by $Y_{s}^{[k]} \to Y_{s}^{[k+2]}$, we find from 
(\ref{cr3}) that
\begin{eqnarray}
 c^{+}_{k+1+l,-k-2+l}  \Eqn{=} \prod_{l=0}^{k}  \bigl(Y_{2k+1-2l}^{[2l-1]}\bigr)^{(-1)^{l}}
  =T^{[2l-1]}_{2k+2} ,  \nn \\
  c^{+}_{k+l,-k+l}  \Eqn{=} \prod_{l=0}^{\frac{n-3}{2}-k}  \bigl(Y_{2k+2l}^{[2l]}\bigr)^{(-1)^{l}}
=T_{2k-1}^{[2l]} .
\end{eqnarray}
Graphically, the $\bbZ_{n}$-transformation generates rotations 
of the polygons represented by $c_{i,j}^{\pm}$.
The cross-ratios $c_{i,j}^{-}$ are obtained from $c_{i,j}^{+}$ simply by the shift 
$Y_{s}^{[k]} \to Y_{s}^{[k+1]}$. We then  find that the expression  for  the cross-ratios
are further summarized in  the form,
\begin{equation}\label{eq:cij-odd}
c^{+}_{i,j}=T^{[i+j]}_{|i-j|-1}, \quad 
c^{-}_{i,j}=T^{[i+j+1]}_{|i-j|-1}.
\end{equation}

This formula  gives a  concise expression of $\Delta A_{\rm BDS}$
in (\ref{eq:delabdsodd1}). Furthermore, by using the quasi-periodicity (\ref{eq:period-Todd}),
one can derive an expression in terms of $T_{s}$ with $s \leq (n-3)/2$,
\begin{equation}\label{DelABDSTsodd}
\Delta A_{\rm BDS}=-{1\over4}\sum_{s=1}^{\frac{n-3}{2}}\sum_{k=1}^{2n} \log \frac{T_s^{[k-1]}}{T_{s-1}^{[k]}}
\log \frac{T_s^{[k]}}{T_{s-1}^{[k-1]}}
-{1\over4}
\sum_{k=1}^n \log \frac{T_{\frac{n-3}{2}}^{[k-1]}}{T_{\frac{n-3}{2}}^{[k+n]}}
\log \frac{T_{\frac{n-3}{2}}^{[k]}}{T_{\frac{n-3}{2}}^{[k+n-1]}},
\end{equation}
where $T_0=1$. This is used for studying  the expansion of the remainder function 
in the next section.

\subsection{Even $n$ case}

Let us move on to  the case of even $n$.
In this case, the sequential cross-ratios $\hat{c}^{\pm}_{i,j}$ are generally 
more complicated than $c^{\pm}_{i,j}$ for odd $n$. However,  it turns out
that still $\hat{c}_{i,j}^{\pm}$ are concisely represented by the Y-/T-functions:
choosing the range of the indices as $ 1 \leq i,j \leq n+1$, we find that 
\eqb\label{chatT}
  \hatc_{i,j}^{+} \Eqn{=} \left\{
                      \begin{array}{ll}
                         T_{|i-j|-1}^{ [ i+j ] } &( \,  i-j  : \, \mbox{odd} \, )  \\
        T_{|i-j|-1}^{ [ i+j ]}  (T_{1}^{ [2 ]})^{(-1)^{j+1}} 
         (T_{n-2}^{ [ -1]})^{(-1)^{j+\frac{n}{2}}}
                          &  (\, i-j  : \, \mbox{even} \, )
                      \end{array}
                     \right.
                     \comma \nn \\
        \hatc_{i,j}^{-} \Eqn{=} \left\{
                      \begin{array}{ll}
                         T_{|i-j|-1}^{ [ i+j +1] } &( \,  i-j  : \, \mbox{odd} \, )  \\
        T_{|i-j|-1}^{ [ i+j +1]}  (T_{1}^{ [ 3 ]})^{(-1)^{j+1}} 
         (T_{n-2}^{ [ 0]})^{(-1)^{j+\frac{n}{2}}}
                          &  (\, i-j  : \, \mbox{even} \, )
                      \end{array}
                     \right.
                     \comma             
\eqe
for  $2 \leq |i-j| \leq n-1$ and $\hatc_{i,j}^{\pm} =1$ otherwise.
For details, see the appendix.
Since the quasi-periodicity (\ref{eq:period-T}) for even $n$ 
involves the factor of $T_{n-2}$, the expression is modified if we choose 
a different  range of $i,j$. Note also that $(-1)^{j} =  (-1)^{i} $  for even $|i-j|$ 
and the above expression is symmetric with respect to $i$ and $j$.

This formula  gives a  concise expression of $\Delta A_{\rm BDS}$ for even $n$
in (\ref{DelABDSeven1}). Furthermore, similarly to the case of odd $n$, 
one finds an expression in terms of $T_{s}$ with $s \leq (n-2)/2$
by using the quasi-periodicity (\ref{eq:period-T}) :
\begin{align}\label{DelABDS}
\Delta A_{\rm BDS}&=-\frac{1}{4}\sum_{s=1}^{\frac{n}{2}-1}
\sum_{k=1}^{2n}\log \frac{T_s^{[k-1]}}{T_{s-1}^{[k]}}
\log  \frac{T_s^{[k]}}{T_{s-1}^{[k-1]}}
+\frac{1}{2}\sum_{k=0}^n \log T_{\frac{n}{2}-1}^{[k]} \log T_{n-2}^{[k+\frac{n}{2}+2]} \notag \\
&\hspace{0.5cm}-\frac{n-1}{4}\log T_{n-2}^{[0]} \log T_{n-2}^{[-1]}
+\frac{(-1)^{\frac{n}{2}+1}}{2}\log T_{n-2}^{[-1]} \log T_1^{[3]}\\
&\hspace{0.5cm}
+\frac{(-1)^{\frac{n}{2}}}{2}\log T_{n-2}^{[0]}
\log \Biggl[ T_1^{[2]} \prod_{l=1}^{[\frac{n}{4}]} (T_{2l-1}^{[0]})^{2(-1)^l} \Biggr]
 \notag 
\end{align}
where $T_0=1$, and $[n/4]$ in the product stands for the greatest integer less 
than or equal to $n/4$ (Gauss symbol).
Since $T_{n-2}(\theta)=e^{-\nu_{n-2} \cosh \theta}$ for real masses, 
$ \log T_{n-2}^{[-1]}$ 
and hence the middle line in (\ref{DelABDS}) 
vanish in this case.

\subsection{Remainder function}

Given the expression of $\Delta A_{\rm BDS}$ in terms of the T-functions, 
the remainder function for the $2n$-point amplitudes at strong coupling 
is summarized as  follows:
For odd $n$, 
\begin{eqnarray}
R_{2n} \Eqn{=} {7\pi\over 12}(n-2)+{1\over2\pi}\int_{-\infty}^{\infty}d\theta
\sum_{s=1}^{n-3}|m_s| \cosh\theta \log\bigl(1+\tilde{Y}_{s}(\theta) \bigr)
\nonumber\\
&&-{1\over4}\sum_{k=1}^{(n-3)/2}
\sum_{j=k}^{(n-3)/2}(-1)^{j+k}(m_{2j}\bar{m}_{2k-1}+\bar{m}_{2j}m_{2k-1})
\nonumber\\
&&
-{1\over4}\sum_{s=1}^{(n-3)/2}\sum_{k=1}^{2n} \log \frac{T_s^{[k-1]}}{T_{s-1}^{[k]}}
\log \frac{T_s^{[k]}}{T_{s-1}^{[k-1]}}
-{1\over4}
\sum_{k=1}^n \log \frac{T_{(n-3)/2}^{[k-1]}}{T_{(n-3)/2}^{[k+n]}}
\log \frac{T_{(n-3)/2}^{[k]}}{T_{(n-3)/2}^{[k+n-1]}}, 
\label{eq:R-odd2}
\end{eqnarray}
For even $n$, $A_{\rm extra}$ and $\Delta A_{\rm BDS}$ add up to be simplified,
and give 
\begin{eqnarray}
R_{2n}  \Eqn{=} {7\pi\over 12}(n-2)+{1\over2\pi}\int_{-\infty}^{\infty}d\theta
\sum_{s=1}^{n-3}|m_s| \cosh\theta \log\bigl(1+\tilde{Y}_{s}(\theta) \bigr)
\nonumber\\
&&-{1\over4}\sum_{k=2}^{(n-2)/2}
\sum_{j=k}^{(n-2)/2}(-1)^{j+k}(m_{2k-2}\bar{m}_{2j-1}+\bar{m}_{2k-2}m_{2j-1})
\nonumber\\
&&
-\frac{1}{4}\sum_{s=1}^{\frac{n}{2}-1}\sum_{k=1}^{2n}\log \frac{T_s^{[k-1]}}{T_{s-1}^{[k]}}
\log  \frac{T_s^{[k]}}{T_{s-1}^{[k-1]}}
+\frac{1}{2}\sum_{k=0}^n \log T_{\frac{n}{2}-1}^{[k]} \log T_{n-2}^{[k+\frac{n}{2}+2]} \notag \\
& & 
-\frac{n-1}{4}\log T_{n-2}^{[0]} \log T_{n-2}^{[-1]}
+\frac{(-1)^{\frac{n}{2}}}{2}\log T_{n-2}^{[0]}
\log \Biggl[\prod_{l=1}^{[\frac{n}{4}]} (T_{2l-1}^{[0]})^{2(-1)^l} \Biggr].
\label{eq:R-even2}
\end{eqnarray}

Now the  remainder function is written completely in term of the 
T-/Y-functions. By using these expressions and 
the conformal perturbation theory of the underlying integrable models,
we discuss analytic expansions of the remainder function
around the CFT/small-mass limit  in the next section .

\setcounter{equation}{0}
\section{High-temperature expansion}

As noted in section 2,  the TBA equations \eqref{TBA} are identical 
to those of  the homogeneous sine-Gordon model  associated with  
SU($n-2$)$_2/$U($1$)$^{n-3}$.
This HSG model is obtained as an integrable perturbation of 
the coset SU($n-2$)$_2/$U($1$)$^{n-3}$ CFT,
\begin{align}
S=S_{\rm CFT}+\lambda \int\! d^2 x\, \Phi_{\boldsymbol{\lambda},\bar{\boldsymbol{\lambda}}},
\end{align}
where $\Phi_{\boldsymbol{\lambda},\bar{\boldsymbol{\lambda}}}$ 
is the perturbing operator, which is given by a linear combination of 
the weight 0 adjoint operators in the coset CFT.
The coupling constant $\lambda$ is related to the overall mass scale $M$ as
\begin{align}\label{lambda}
\lambda=-\kappa_n M^{2-(\Delta+\bar{\Delta})} ,
\end{align}
where $\Delta=\bar{\Delta}=(n-2)/n$ are the conformal dimensions of 
$\Phi_{\boldsymbol{\lambda},\bar{\boldsymbol{\lambda}}}$ 
and $\kappa_{n}$ is the dimensionless coupling.
In the small-mass limit, one can perturbatively expand 
the physical quantities around the CFT point 
($\lambda = 0$) by using the conformal perturbation theory.
Since the mass scale is proportional to the inverse temperature, 
we call it the high-temperature/small-mass expansion. 
In \cite{Hatsuda:2011ke}, we discussed the high-temperature
expansion in the HSG model. In particular, the Y-/T-functions are expanded 
by using the relation to the $g$-function (boundary entropy)  \cite{Affleck:1991tk}.
Together with the expansion of the free energy, 
we obtained the high-temperature expansion of the remainder 
function at strong coupling for the octagon and for the decagon explicitly.

Here we consider the high-temperature expansion of the remainder function 
for the general $2n$-gon  at strong coupling.
Below we mainly focus on the case that all the masses are real.
The results in the general case
of complex masses are obtained  by complexifying the masses in the
final expression \cite{Hatsuda:2011ke}. The way of the complexification is 
specified by consideration based on the $\bbZ_{2n}$-symmetry (\ref{Z2n}), 
which is equivalent to $m_{s} \to m_{s}/i $ in the high-temperature expansion.

\subsection{Expansion of T-functions}
First, let us consider the expansion of the T-function.
From the periodicity, the Y- and T-functions have the Laurent expansion 
for $|\theta|<\infty$. Each coefficient of the Laurent expansion is further 
 expanded by the scale parameter $l=ML$
near the high-temperature limit, where $L$ is the size of the system. 
See \cite{Hatsuda:2011ke} for detail.
In our notation, the mass $M_j$ of the $j$-th particle is related to $m_j$ 
in the TBA equations \eqref{TBA} as follows,
\begin{align}
|m_j|=M_j L=\tilde{M}_j l,
\end{align}
where $\tilde{M}_j \equiv M_j/M$ is the relative mass.

For odd $n$ case, since the T-functions satisfy the half-periodicity \eqref{eq:period-Todd}, 
the T-functions are expanded as
\begin{align}\label{Texpand}
T_s(\theta)=\sum_{p,q=0}^\infty t_s^{(p,2q)}l^{(1-\Delta)(p+2q)}\cosh \left( \frac{2p\theta}{n} \right),
\end{align}
 with $t_{n-2-s}^{(p,2q)} = (-1)^{p}t_{s}^{(p,2q)}$.
Some of the coefficients $t_s^{(p,2q)}$ are fixed by the T-system.
For example, one can check that
\begin{align}
t_s^{(0,0)}=\frac{\sin(\frac{(s+1)\pi}{n})}{\sin (\frac{\pi}{n})},
\label{eq:ts00}
\end{align}
and
\begin{align}
t_s^{(1,0)}=t_s^{(0,2)}=t_s^{(1,2)}=0.
\label{eq:ts10}
\end{align}
Note that $t_s^{(0,0)}$ is equal to the quantum dimensions (ratios of the modular 
S-matrices) because the T-system reduces to the Q-system at this order. 
One can also check that \eqref{eq:ts00} and \eqref{eq:ts10} are consistent
 with the results from the CFT perturbation.
In \cite{Hatsuda:2011ke}, we determined the first non-trivial coefficient $t_s^{(2,0)}$ as
\begin{align}
\frac{t_s^{(2,0)}}{t_{s}^{(0,0)}}
=-\frac{\kappa_{n}G(\tilde{M}_j)B(1-2\Delta,\Delta)}{2(2\pi)^{1-2\Delta}}
\( \frac{\sin(\frac{3(s+1)\pi}{n})}{\sin(\frac{(s+1)\pi}{n})} 
\sqrt{\frac{\sin(\frac{\pi}{n})}{\sin(\frac{3\pi}{n})}}-
\sqrt{\frac{\sin(\frac{3\pi}{n})}{\sin(\frac{\pi}{n})}}\),
\label{eq:ts20}
\end{align}
where $B(x,y)\equiv \Gamma(x)\Gamma(y)/\Gamma(x+y)$ is the beta function, and
$G(\tilde{M}_j)$ is the normalization factor of the two-point function of 
the perturbing operator $\Phi_{\boldsymbol{\lambda},\bar{\boldsymbol{\lambda}}}$,
\begin{align}\label{Phi2pt}
\Bigl\langle \Phi_{\boldsymbol{\lambda},\bar{\boldsymbol{\lambda}}}(z)
 \Phi_{\boldsymbol{\lambda},\bar{\boldsymbol{\lambda}}}(0) \Bigr\rangle
= \frac{G^{2}}{ |z|^{4\Delta}  } , \qquad
  G(\tilde{M}_j) = \sum_{i,j=1}^{n-3} \tilde{M}_i^{\frac{2}{n}} F_{ij} \tilde{M}_j^{\frac{2}{n}}.
\end{align}  
Classically, the coefficients $F_{ij}$
are given by the inverse of the Cartan matrix.
At the quantum level, however, 
these coefficients receive corrections due to the renormalization.

For even $n$ case, the expansion is slightly complicated due to the extra factor 
in the quasi-periodicity~\eqref{eq:period-T}.
From the quasi-periodicity, we find that the T-functions should 
have the following forms,\footnote{
The quasi-periodicity constrains the form of the exponent up to 
$(\theta -c_{k}) \sinh \theta$, where $c_{k}$ is a constant.
For real masses, the reality condition $\overline{T(\theta)} = T(-\bar{\theta})$
requires $c_{k} =0$. For $n=4$ with complex masses,
this constant is precisely the phase of the mass parameter. 
One can also check for lower $n$
that $c_{k}$ is independent of $k$ to satisfy the T-system.
}
\begin{alignat}{3} 
T_{2k+1}(\theta)&=\hat{T}_{2k+1}(\theta)\; e^{(-1)^{k+\frac{n}{2}}
\frac{2}{n\pi} \nu_{n-2} \theta \sinh \theta}
&\quad &\(k=0,\dots, \frac{n}{2}-2\), \nonumber  \\ 
T_{2k}(\theta)&=\hat{T}_{2k}(\theta)\; 
e^{(-1)^{k+\frac{n}{2}}
\frac{1}{2} \nu_{n-2} \cosh \theta}
& &\(k=1,\dots,\frac{n}{2}-2\).  \label{Teven2}
\end{alignat}
 Here,  $\hat{T}_s(\theta)$ satisfy the half-periodicity
$\hat{T}_s(\theta+\pi i n/2)=\hat{T}_{n-2-s}(\theta)$,
and are expanded near  the high-temperature limit as
\begin{align}\label{That}
\hat{T}_s(\theta)=\sum_{p,q=0}^\infty \hat{t}_s^{(p,2q)}
l^{(1-\Delta)(p+2q)}\cosh \left( \frac{2p\theta}{n} \right).
\end{align}
Since the exponential factors in (\ref{Teven2}) start to be non-trivial
at ${\cal O} (l)$, the coefficients 
$\hat{t}_s^{(p,2q)}$ for lower $p$ and $q$ are the same as $t_s^{(p,2q)}$.
Consequently, we have the same formulae \eqref{eq:ts00} 
and \eqref{eq:ts10} as in the odd $n$ case, and 
$ \hat{t}_{s}^{(2,0)} = t_{s}^{(2,0)} $ are given by (\ref{eq:ts20}) 
for $n \geq 6$.\footnote{
For $n=4$, $ \hat{t}_{1}^{(2,0)}$ differs from  $ t_{1}^{(2,0)}$ due to the factor 
in (\ref{Teven2}), and (\ref{eq:ts20}) does not make sense because $\Delta = 1/2$.
The relation between the T-function and the $g$-function, from which  (\ref{eq:ts20})
is derived, is based on the integral equations obeyed 
by them and not on the particular form of the expansion.
One can numerically check (\ref{eq:ts20}) for even $n$, 
as was done for odd $n$ \cite{Hatsuda:2011ke}.
}
We also note  that $T_{2k+1}(\theta)$ for even $n$ contain 
the non-analytic term $\log l$ in the high-temperature expansion.
This follows from the integral equations (\ref{integraleqT}) and 
the above expansions (\ref{Teven2}). The non-analytic terms are
cancelled in $Y_{s}$ and $T_{2k}$, which are given by the ratios of $Y_{s}$. 
We will come to this point later again. 

\subsection{Expansion of remainder function at strong coupling} 
Now let us consider the high-temperature expansion of the remainder function 
at strong coupling. As seen in the previous sections, the remainder function is given 
by \eqref{eq:R-odd} or \eqref{eq:R-odd2} 
for odd $n$ and by \eqref{eq:R-even} or \eqref{eq:R-even2} for even $n$.

\subsubsection{Odd $n$ case}
Let us first consider the odd $n$ case.
In this case, the period term is given by (\ref{Aperiodsodd}).
As seen in \cite{Hatsuda:2011ke}, the CFT perturbation allows us to expand the free energy 
part as
\begin{align}
A_{\rm free}=\frac{\pi}{6}c_n +f_n^{\rm bulk}
+\sum_{k=2}^\infty f_n^{(k)}l^{\frac{4k}{n}},
\label{eq:exp-Afree}
\end{align}
where $c_n$ is the central charge of the coset CFT 
for SU($n-2$)$_2/$U($1$)$^{n-3}$, and 
$f_n^{\rm bulk}$ is the bulk contribution.  They  are given respectively by
\begin{align}
c_n=\frac{(n-2)(n-3)}{n},\quad f_n^{\rm bulk}
=\frac{1}{4}\sum_{j,k=1}^{n-3}m_j (I^{-1})_{jk} m_k,
\end{align}
with $I_{jk}$ being the incidence matrix for $A_{n-3}$.
One can check that the bulk term $f_n^{\rm bulk}$ just cancels 
with the period part $A_{\rm periods}$.%
\footnote{
Useful relations to see this are  
$Z_{j} = -\frac{1}{2} \eta_{jk}m_{k}$ and the one
between the incidence matrix and the intersection matrix
$I =  i \eta \theta  \eta^{-1}$ 
where $\eta =$diag$(i,1,i, \cdots )$.
}
The corrections $f_n^{(k)}$ in \eqref{eq:exp-Afree} are given 
by the worldsheet integral of the connected $k$-point function of the perturbing operator.
For $k=2$, we have
\begin{align}
f_{n}^{(2)} = \frac{\pi}{6} C_{n}^{(2)} \kappa_{n}^{2}G^{2}(\tilde{M}_{j}) , 
\label{eq:fn2}
\end{align}
where 
\begin{align}
C_{n}^{(2)} = 3 (2\pi)^{\frac{2(n-4)}{n}} 
  \gamma^{2}\(\frac{n-2}{n}\)\gamma\(\frac{4-n}{n}\),
\end{align}
and $\gamma(x)\equiv \Gamma(x)/\Gamma(1-x)$.

Since $\Delta A_{\rm BDS}$ is expressed in terms of the T-functions as in 
\eqref{DelABDSTsodd}, we find the high-temperature expansion
of $\Delta A_{\rm BDS}$ after substituting \eqref{Texpand} 
into \eqref{DelABDSTsodd},
\begin{align}
\Delta A_{\rm BDS}=-\frac{n}{2}\sum_{s=1}^{(n-3)/2} 
\log^2 \(\frac{t_s^{(0,0)}}{t_{s-1}^{(0,0)}}\)
-l^{\frac{8}{n}}\frac{n}{4}\left[
\sum_{s=1}^{(n-3)/2} A_{n,s}-2
\( \frac{t_{\frac{n-3}{2}}^{(2,0)}}{t_{\frac{n-3}{2}}^{(0,0)}}\)^2
\sin^2\(\frac{\pi}{n}\)\right] , 
\end{align} 
where
\begin{align}
A_{n,s} &\equiv \left[ \( \frac{t_{s-1}^{(2,0)}}{t_{s-1}^{(0,0)}}\)^2+\( \frac{t_{s}^{(2,0)}}{t_{s}^{(0,0)}}\)^2 \right]
\cos\(\frac{2\pi}{n} \)
-\frac{2t_{s-1}^{(2,0)}t_s^{(2,0)}}{t_{s-1}^{(0,0)}t_{s}^{(0,0)}} \nonumber  \\
&\hspace{0.5cm}+\left[ \( \frac{t_{s-1}^{(2,0)}}{t_{s-1}^{(0,0)}}\)^2
-\( \frac{t_{s}^{(2,0)}}{t_{s}^{(0,0)}}\)^2-
4\(\frac{t_{s-1}^{(0,4)}}{t_{s-1}^{(0,0)}}-\frac{t_{s}^{(0,4)}}{t_{s}^{(0,0)}}\)\right]
\log \( \frac{t_s^{(0,0)}}{t_{s-1}^{(0,0)}} \) ,
\label{eq:Ans}
\end{align}
and we have used \eqref{eq:ts10}.
The coefficients $t_s^{(0,0)}$ and $t_s^{(2,0)}$ are given by \eqref{eq:ts00} 
and \eqref{eq:ts20}, respectively. 
For $t_s^{(0,4)}$, we have the equations which follow from the
T-system,
\begin{align}
2t_s^{(0,0)}t_s^{(0,4)}+\frac{1}{2}(t_s^{(2,0)})^2\cos\( \frac{4\pi}{n} \)
=t_{s-1}^{(0,0)}t_{s+1}^{(0,4)}+t_{s+1}^{(0,0)}t_{s-1}^{(0,4)}
+\frac{1}{2}t_{s-1}^{(2,0)}t_{s+1}^{(2,0)},
\label{eq:ts04}
\end{align}
for $s=1,\dots,n-3$.
By solving these equations with the boundary condition $T_0=T_{n-2}=1$,
the coefficients $t_s^{(0,4)}$ are expressed in terms of $t_s^{(0,0)}$ 
and $ t_s^{(2,0)}$.

We note that $t_{s}^{(3,0)}, t_{s}^{(2,2)}$ and $t_{s}^{(4,0)}$ do not appear in the expansion.
This is understood as a consequence of the $\bbZ_{2n}$-symmetry: 
For general complex $m_{s}$, the terms in the expansion (\ref{Texpand}) are modified 
\cite{Hatsuda:2011ke} as
$t_{s}^{(p,2q)} \cosh (2p \theta/n)\to \frac{1}{2} ( t_{s}^{(p,2q)}  e^{2p \theta/n} 
+ \bar{t}_{s}^{(p,2q)}  e^{-2p \theta/n}) $. 
Under the $\bbZ_{2n}$-transformation (\ref{Z2n}),
these coefficients transform as $(t_{s}^{(p,2q)},  \bar{t}_{s}^{(p,2q)}) 
\to (t_{s}^{(p,2q)} e^{p \pi i/n},  \bar{t}_{s}^{(p,2q)}  e^{-p \pi i/n})$. 
Given the vanishing coefficients (\ref{eq:ts10}) at lower orders, 
the non-constant combinations invariant under the $\bbZ_{2n}$-symmetry 
are only $t_{s}^{{(2,0)}} \bar{t}_{s}^{{(2,0)}} $ and $t_{s}^{(0,4)}$ 
up to ${\cal O}(l^{\frac{8}{n}})$.

Combining all of the above  results, we then find that the remainder function 
at strong coupling has the following high-temperature expansion,
\begin{align}\label{R2nodd1}
R_{2n}=R_{2n}^{(0)}+l^{\frac{8}{n}}R_{2n}^{(4)}+{\cal O}(l^{\frac{12}{n}}),
\end{align}
where
\begin{align}
R_{2n}^{(0)}&=\frac{\pi}{4n}(n-2)(3n-2)-\frac{n}{2}\sum_{s=1}^{(n-3)/2} 
\log^2 \( \frac{\sin (\frac{(s+1)\pi}{n})}{\sin (\frac{s\pi}{n})} \),  \label{R2nodd2} \\
R_{2n}^{(4)}&=\frac{\pi}{6}C_n^{(2)}\kappa_n^2 G^2(\tilde{M}_j)
-\frac{n}{4}\left[
\sum_{s=1}^{(n-3)/2} A_{n,s}-2
\( \frac{t_{\frac{n-3}{2}}^{(2,0)}}{t_{\frac{n-3}{2}}^{(0,0)}}\)^2
\sin^2\(\frac{\pi}{n}\)\right].  \label{R2nodd3}
\end{align}
Note that the leading term $R_{2n}^{(0)}$ gives the remainder function 
for the regular $2n$-gon.

By further using  (\ref{eq:ts20}) and (\ref{eq:ts04}),
the results are expressed by $t_{s}^{(0,0)}$ and, e.g., $t_{1}^{(2,0)}$.
All the mass parameter dependence is encoded in the latter.
The results for complex masses are given by replacing $(t_{1}^{(2,0)})^{2}$ 
in the resultant expression by $t_{1}^{(2,0)} \bar{t}_{1}^{(2,0)}$. 
One can also express the result in terms of the expansion coefficients 
of the Y-function $y_{s}^{(2,0)}$,
which are defined similarly to $t_{s}^{(2,0)}$, 
by using the relation,
\eqb\label{y20t20}
 y_{s}^{(2,0)} = 2 \cos \Bigl(\frac{2\pi}{n} \Bigr) t_{s}^{(0,0)} t_{s}^{(2,0)} 
 \quad  \ (n \geq 5) .
\eqe

\subsubsection{Even $n$ case}
Let us next consider the even $n$ case.
In this case, the period term is given by (\ref{Aperiodseven}).
The free energy part is expanded as in \eqref{eq:exp-Afree}, but the bulk term 
is now given by
\cite{Hatsuda:2011ke} 
\begin{align}
f_n^{\rm bulk}=\frac{1}{n \pi} \nu_{n-2}^2 \log l.
\end{align}
As in \eqref{eq:Aextra}, $A_{\rm extra}$ is expressed by $T_1$ and $T_{n-2}$.
It contains the non-analytic term $\log l$ coming from $T_{1}$, and 
this  is canceled by $f_{n}^{\rm bulk}$.
To see this, let us recall that $L_1 \equiv \log(1+Y_1)$ has the following order 
$l$ term \cite{Zamolodchikov:1991vx}
\begin{align}
L_1 \sim -\frac{2}{n}\nu_{n-2} \cosh\( \theta +\frac{in\pi}{2} \)
=(-1)^{\frac{n}{2}+1}\frac{2}{n}\nu_{n-2}\cosh \theta.
\end{align}
This term leads to the $\log l$ term in $\log T_1$ as
\begin{align}
\log T_1(\theta_0) 
 \sim  \int_{-\log(1/l)}^{\log(1/l)} \frac{d\theta}{2\pi} 
\frac{L_1(\theta)}{\cosh(\theta_0-\theta)}
 \sim(-1)^{\frac{n}{2}}\frac{2}{n\pi}\nu_{n-2}\cosh \theta_0  \log l ,
\end{align}
where $\theta_0$ is a constant.
Since $A_{\rm extra}$ reduces to
\begin{align}
A_{\rm extra}=\frac{(-1)^{\frac{n}{2}+1}}{2}\log T_{n-2}(0) \log T_1(\pi i) ,
\end{align}
for  real masses, we have
\begin{align}
A_{\rm extra} \sim -\frac{1}{n\pi}\nu_{n-2}^2 \log l+{\cal O}(l),
\end{align}
which indeed cancels $f_n^{\rm bulk}$. We also note that
the analytic term of $A_{\rm extra}$ starts from order $l$, 
because $\log T_{n-2}(0)=-\nu_{n-2}$  is of  order $l$.

For the expansion of $\Delta A_{\rm BDS}$,  we first note that 
\begin{align}
\log T_s(\theta)=\log \hat{T}_s(\theta)+{\cal O}(l),
\end{align} 
and  the exponential factors in (\ref{Teven2}) and $T_{n-2}$
are irrelevant up to ${\cal O}(l^{\frac{8}{n}})$ for $ n \geq 10$.
Thus, similarly to the odd $n$ case, 
$\Delta A_{\rm BDS}$ for $n\geq 10$ is expanded as%
\footnote{As we have mentioned, $\log T_{2k+1}$ 
contain the non-analytic terms. Such terms, however, do not appear 
in $\Delta A_{\rm BDS}$, because $\Delta A_{\rm BDS}$ 
is originally expressed by the cross-ratios $\hat{c}_{i,j}^{\pm}$ and 
these cross-ratios can be expressed by the Y-functions only, which
do not have the non-analytic terms.
} 
\begin{align}
\Delta A_{\rm BDS}
=-\frac{n}{2}\sum_{s=1}^{n/2-1} \log^2 \( \frac{t_s^{(0,0)}}{t_{s-1}^{(0,0)}}\)
-l^{\frac{8}{n}}\frac{n}{4}
\sum_{s=1}^{n/2-1} 
\hat{A}_{n,s} + {\cal O}(l^{\frac{12}{n}}) ,
\end{align}
where $\hat{A}_{n,s}$ is given by replacing $t_s^{(p,2q)}$ in $\eqref{eq:Ans}$ 
by $\hat{t}_s^{(p,2q)}$. These are, however, given by (\ref{eq:ts00}), 
(\ref{eq:ts20}) and (\ref{eq:ts04}) as in the case of odd $n$, 
since $t_{s}^{(p,2q)} = \hat{t}_{s}^{(p,2q)}$ 
for $p+2q < n/2$.
Combining the relevant terms from $A_{\rm free}$ and $\Delta A_{\rm BDS}$, 
we find for $n \geq 10$ that 
\begin{align}
R_{2n}=R_{2n}^{(0)}+l^{\frac{8}{n}}R_{2n}^{(4)}+{\cal O}(l^{\frac{12}{n}}),
\label{eq:R2neven1}
\end{align}
where
\begin{align}
R_{2n}^{(0)}&=\frac{\pi}{4n}(n-2)(3n-2)-\frac{n}{2}\sum_{s=1}^{n/2-1} 
\log^2 \( \frac{\sin (\frac{(s+1)\pi}{n})}{\sin (\frac{s\pi}{n})} \), 
\label{eq:R0-even}\\
R_{2n}^{(4)}&=\frac{\pi}{6}C_n^{(2)}\kappa_n^2 G^2(\tilde{M}_j)
-\frac{n}{4}
\sum_{s=1}^{n/2-1} \hat{A}_{n,s}.
\label{eq:R2-even}
\end{align}

For $n=4$, we can obtain the all order expansion in $l$. 
The logarithmic terms there are canceled out,
and the remainder function is expanded in $l^{2}$. 
See \cite{Hatsuda:2011ke} for detail.
We have also checked that the final results \eqref{eq:R2neven1}-\eqref{eq:R2-even} 
are valid for $n=6,8$: the contributions from the extra factors in the T-functions in 
(\ref{Teven2}) exactly cancel with those from $A_{\rm extra}$, so that 
the final result becomes $\bbZ_{2n}$-symmetric.
The relations (\ref{eq:ts00}), (\ref{eq:ts10}),  (\ref{eq:ts20}) and
(\ref{eq:ts04}) also hold.

As in the case of odd $n$, the results for complex masses 
are given by expressing $t_{s}^{(2,0)}$ and $ \kappa_{n}G$, e.g., by $t_{1}^{(2,0)}$ and 
replacing $(t_{1}^{(2,0)})^{2}$  by $t_{1}^{(2,0)} \bar{t}_{1}^{(2,0)}$ (for $n \geq 6 $).
In Table 1 of section 6, we list the numerical values of $R_{2n}^{(0)}$ 
and $R_{2n}^{(4)}/t_{1}^{(2,0)} \bar{t}_{1}^{(2,0)}$ for both odd and even $n$.
Since the mass-parameter dependence is encoded in  
$t_{1}^{(2,0)} \bar{t}_{1}^{(2,0)}$, they are independent of 
$m_{s}$ and $\kappa_{n} G$.

\subsection{Relation between cross-ratios and mass parameters}

In summary, the leading correction to the remainder function
around the CFT limit is expressed by the coefficients $t_{s}^{(0,0)}$, $t^{(2,0)}_s$ 
and $\kappa_n G$, where $\kappa_n$ is the dimensionless coupling 
defined in (\ref{lambda}) and  $G$ is the normalization 
factor of the two-point function in (\ref{Phi2pt}).
From (\ref{eq:ts20}),  $t_{s}^{(2,0)}$ are also regarded as functions of  $\kappa_{n}G$,
and vice versa. Thus, the result is given in terms of (one of) $t_{s}^{(2,0)}$ or $\kappa_{n} G$,
which are functions of the mass parameters.

The momentum dependence  of 
the remainder function is read off through 
their relation to the Y-functions. 
Indeed, similarly to $T_{s}$ the Y-functions  for general complex masses  
are expanded  near the CFT limit as
\eqb
   Y_{s}(\theta) = y_{s}^{(0,0)} + \frac{1}{2} \Bigl( y_{s}^{(2,0)} e^{{4 \over n} \theta} 
   + \bar{y}_{s}^{(0,0)} e^{- {4 \over n} \theta} \Bigr) l^{\frac{4}{n}}  
      + {\cal O}(l^{\frac{6}{n}}).
\eqe
Here, $y_{s}^{(0,0)}$ are the solution to the constant Y-system, 
$y_{s}^{(0,0)} = \sin \bigl( \frac{s\pi}{n} \bigr) \bigl( \frac{(s+2)\pi}{n} \bigr)/ 
\sin^{2} \bigl( \frac{\pi}{n} \bigr)$. For real $m_{s}$, one has
$y_{s}^{(2,0)} = \bar{y}_{s}^{(2,0)}$.
From (\ref{y20t20}), we then find that
\eqb \label{eq:Ysk}
   Y_{s}^{[k]} = y_{s}^{(0,0)}  + \cos\Bigl( \frac{2\pi}{n} \Bigr) t_{s}^{(0,0)}
    \Bigl( t_{s}^{(2,0)} e^{{2\pi \over n} ki } + 
      \bar{t}_{s}^{(2,0)} e^{- {2\pi \over n} ki} \Bigr) l^{\frac{4}{n}}  
      + {\cal O}(l^{\frac{6}{n}}) .
\eqe
By inverting these relations, the remainder function is 
expressed in terms of the cross-ratios of momenta
(which depend on each other at this order through (\ref{eq:Ysk})). 
To be concrete, one finds that
\eqb\label{tvarphi}
   |t_{s}^{(2,0)}| l^{\frac{4}{n}} \Eqn{=} \frac{\delta Y_{s}^{[0]}}{2 t_{s}^{(0,0)} 
   \cos \frac{2\pi}{n} \cos \varphi_{s}}
   \comma \nn \\
   \frac{2\pi}{n} \varphi_{s} \Eqn{=} \arctan\biggl( \cot\Bigl(\frac{2\pi}{n} \Bigr)
   \frac{\delta Y_{s}^{[-1]} - \delta Y_{s}^{[1]}}{\delta Y_{s}^{[-1]} + \delta Y_{s}^{[1]}} \biggr)
   \comma 
\eqe
at this order. Here, we have set $t_{s}^{(2,0)} 
= |t_{s}^{(2,0)}| e^{i\varphi_{s}}$, and  
$\delta Y_{s}^{[k]}$ are the deviations of the cross-ratios
from the regular-polygonal/small-mass limit,  
$ \delta Y_{s}^{[k]} \equiv Y_{s}^{[k]} - y_{s}^{(0,0)}$.
The cross-ratios are given by (\ref{bracketsChi}), (\ref{bracketsY}) and the 
$\bbZ_{n}$ symmetry (\ref{Zn}) or $Y_{s}^{[k]} \to Y_{s}^{[k+2]}$.
From (\ref{R2nodd3}) and (\ref{eq:R2-even}), 
it then follows that the remainder function depends on these cross-ratios
through $\kappa_{n}^{2} G \bar{G} \ \propto \ | t_{s}^{(2,0)}|^{2} $
given via (\ref{tvarphi}). 
Once $t_{s}^{(2,0)}$ are expressed by $m_{s}$,  
one can also find the momentum dependence  along the trajectories parametrized
by them.
The relation between   the sequential cross-ratios 
$c^{\pm}_{i,j}, \hat{c}_{i,j}^{\pm}$ and $t_{s}^{(2,0)}$ or $\kappa_{n}G$ 
is similarly found from the expansion of $T_{s}$.

\setcounter{equation}{0}
\section{Mass-coupling relations in single mass cases}

 As mentioned at the end of the last section, the momentum dependence
of the leading expansion  
of the remainder function is traced through $t_{s}^{(2,0)}$ or $\kappa_{n} G$
by expressing them as functions of the $n-3$ mass parameters $m_{s}$.
In this section, we find the exact form of $\kappa_{n} G$ and hence of $t_{s}^{(2,0)}$ 
in simple cases 
where the TBA system has only one mass scale.
Such TBA systems associated with various Dynkin diagrams are classified 
in \cite{Ravanini:1992fi},  and our TBA system with a single mass  scale reduces
to the known ones in the classification \cite{Hatsuda:2011ke}. 
We read off $\kappa_{n} G$ thereof. 

The single mass cases discussed below fix   
$(n-1)/2$ parameters for odd $n$ and $(n-2)/2$ for even $n$ among 
the $(n-1)(n-3)/4$ independent parameters in $\kappa_{n} G$ for odd $n$ 
and $(n-2)^2/4$ for even $n$.%
\footnote{There 
are two symmetries for $F_{ij}$, $F_{ij}=F_{ji}$ and $F_{ij}=F_{n-2-i,n-2-j}$, and
one can absorb the over all scale into $l$.}
For $n=5$, these single mass cases completely fix the form of $\kappa_{n} G$ 
\cite{Hatsuda:2011ke}.

\subsection{Case of perturbed unitary minimal model}
Let us first consider the case that only the leftmost mass parameter is non-zero:
\begin{align}
M_1=M,\quad M_2=\dots=M_{n-3}=0.
\end{align}
In this case, the TBA equations for the homogeneous sine-Gordon theory reduces to
those for the (RSOS)$_{n-2}$ scattering theory \cite{Zamolodchikov:1991vh,Itoyama:1990pv},
 which is regarded as the massive
perturbation of the unitary minimal model ${\cal M}_{n-1,n}$ by the primary field $\Phi_{1,3}$.
Taking into account an appropriate normalization of the overall scale,
we then find
\begin{align}
\kappa_n G(\tilde{M}_j) = \kappa_n F_{11} = \kappa_n^{\rm RSOS},
\label{eq:norm-RSOS}
\end{align} 
where the constant $\kappa_n^{\rm RSOS}$ is given  by  \cite{Zamolodchikov:1991vh} 
\begin{align}
\kappa_n^{\rm RSOS}=
\frac{1}{\pi} \frac{n^{2}}{(n-2)(2n-3)} 
   \biggl[ \gamma\Bigl( \frac{3(n-1)}{n}\Bigr) \gamma\Bigl( \frac{n-1}{n}\Bigr)
   \biggr]^{\frac{1}{2}}
   \biggl[ \frac{\sqrt{\pi}\Gamma(\frac{n}{2})}{2\Gamma(\frac{n-1}{2})}
   \biggr]^{ \frac{4}{n} }.
   \label{eq:kappaRSOS}
\end{align}
Although
\eqref{eq:norm-RSOS} and \eqref{eq:kappaRSOS} have already been given
in \cite{Hatsuda:2011ke}, we have included them for completeness.
Given this coupling, we also find that the first non-trivial 
coefficient \eqref{eq:fn2} becomes,
\begin{align}
f_n^{(2)}=2\pi \( \frac{n-3}{n-2} \)^2  
\biggl[ \frac{1}{4\sqrt{\pi}}\frac{\Gamma(\frac{n}{2})}{\Gamma(\frac{n-1}{2})} \biggr]^{\frac{8}{n}}
\gamma\(\frac{4}{n}-1\)\gamma\( 1-\frac{3}{n}\)
\gamma^2\( 1-\frac{2}{n}\) \gamma\(1-\frac{1}{n}\). 
\end{align}

\subsection{Case of perturbed unitary SU(2) diagonal coset model}

Let us next consider the case that only the $k$-th mass parameter is
non-zero:%
 \footnote{
The result in this subsection is based on discussions with Kazuhiro Sakai.}
\begin{align}
M_j=\delta_{jk}M \;\;{\rm for}\;\; j=1,\dots,n-3.
\label{eq:kth-mass}
\end{align}
In the following, we take the normalization,
\begin{align}
\kappa_n G(\tilde{M}_j)=\kappa_n^{\rm RSOS} \sum_{i,j=1}^{n-3}
\tilde{M}_i^\frac{2}{n} \frac{F_{ij}}{F_{11}} \tilde{M}_j^\frac{2}{n},
\label{eq:knG}
\end{align}
which reduces to  \eqref{eq:norm-RSOS} in the previous case, and 
\begin{align}
\kappa_n G(\tilde{M}_j) = \kappa_n^{\rm RSOS} \frac{F_{kk}}{F_{11}},
\label{eq:knG-2}
\end{align}
in the more general present case.

The TBA equations \eqref{TBA} with the real masses \eqref{eq:kth-mass} 
describe the system obtained as the integrable perturbation of
the ${\rm SU}(2)_{k} \times {\rm SU}(2)_{n-2-k} / {\rm SU}(2)_{n-2}$
coset CFT  by the operator 
$\Phi_{(\mathbf{1},\mathbf{1};\textbf{adj})}$ \cite{Zamolodchikov:1991vg}. 
In \cite{Fateev:1993av}, the exact mass-coupling relation and the high-temperature expansion
of the free energy in the perturbed coset $G_k \times G_l/G_{k+l}$  theories have been given.  
Applying this result to our case of $G={\rm SU}(2)$, one obtains 
\begin{align}
f_n^{(2)}&=2\pi \frac{k^2(n-k-2)^2}{(n-2)^2}
\left[ \frac{1}{8} \frac{\Gamma(\frac{n}{2})}{\Gamma(\frac{k}{2}+1)\Gamma(\frac{n-k}{2})}
\right]^{\frac{8}{n}} \nonumber \\
& \qquad \quad \times \
\gamma\(\frac{4}{n}-1\)\gamma\( 1-\frac{3}{n}\)
\gamma^2\( 1-\frac{2}{n}\) \gamma\(1-\frac{1}{n}\).
\end{align}
On the other hand, from \eqref{eq:knG-2}, we find
\begin{align}
f_n^{(2)}=\frac{\pi}{6}C_n^{(2)}(\kappa_n^{\rm RSOS})^2 \(\frac{F_{kk}}{F_{11}}\)^2.
\label{eq:free-kth}
\end{align}
Comparing these two expressions, we can fix the unknown ratio $F_{kk}/F_{11}$ 
in $\kappa_{n} G$ as 
\begin{align}
\frac{F_{kk}}{F_{11}}=\frac{k(n-k-2)}{n-3}
\left[ \frac{\sqrt{\pi}}{2} \frac{\Gamma(\frac{n-1}{2})}{\Gamma(\frac{k}{2}+1)\Gamma(\frac{n-k}{2})}
\right]^{\frac{4}{n}}.
\end{align}

\subsection{Case of perturbed non-unitary minimal model}

When $n$ is odd, we can further consider the case where
\begin{align}
M_1=M_{n-3}=M, \quad M_2=\dots=M_{n-4}=0.
\label{eq:type-T}
\end{align}
In this case,  with the normalization \eqref{eq:knG} we have 
\begin{align}
\kappa_n G(\tilde{M}_j)= 2\kappa_n^{\rm RSOS}\(1+\frac{F_{1,n-3}}{F_{11}}\),
\label{eq:G-T}
\end{align}
where we have used the symmetries,  $F_{11}=F_{n-3,n-3}$, $F_{1,n-3}=F_{n-3,1}$.
Since the Y-functions satisfy the additional relation $Y_s(\theta)=Y_{n-2-s}(\theta)$,
the number of independent Y-functions reduces to half. 
One then  finds that  the resultant reduced TBA system is equivalent to that for the 
$T_{(n-3)/2}=A_{n-2}/\mathbb{Z}_2$ scattering theory,  
which is described by the perturbation of the non-unitary coset 
${\rm SU}(2)_{n/2-3} \times {\rm SU}(2)_1 / {\rm SU}(2)_{n/2-2}$ 
model by $\Phi_{(\mathbf{1},\mathbf{1};\textbf{adj})}$,
or equivalently of the non-unitary minimal model ${\cal M}_{n-2,n}$ by $\Phi_{1,3}$
\cite{Ravanini:1992fi}. 
Due to the above $\mathbb{Z}_2$-symmetry, the free energy  
for the TBA system \eqref{TBA} with \eqref{eq:type-T} 
is twice larger than that in the perturbed minimal model ${\cal M}_{n-2,n}$ up to 
the constant part corresponding to the central charge.
$\kappa_{n} G $ in \eqref{eq:G-T} is then determined as below.

First, for the $T_{(n-3)/2} $ scattering theory, 
the perturbing operator $\hat{\Phi}=\Phi_{1,3}$ has the dimension
$\hat{\Delta}=(n-4)/n$, 
and the  exact mass-coupling relation \cite{Fateev:1993av} is given by
\begin{align}
\hat{\lambda}=\hat{\kappa} M^{8/n},
\end{align}
where
\begin{align}
\hat{\kappa}^2=\frac{1}{\pi^2}\( \frac{n-6}{n-4}\)^2 \gamma\(1-\frac{2}{n}\) \gamma\( 1-\frac{6}{n}\)
\left[ \frac{\sqrt{\pi}}{2} \frac{\Gamma(\frac{n}{4})}{\Gamma(\frac{n}{4}-\frac{1}{2})} \right]^{\frac{16}{n}}.
\end{align}   
The free energy is expanded for the small mass scale as
\begin{align}\label{eq:hatF}
\hat{F}(l)=\frac{\pi}{6}\hat{c}+\hat{B} l^2+\sum_{k=1}^\infty \hat{f}^{(k)} l^{8k/n},
\end{align}
where $\hat{c}$ is the effective central charge, and $\hat{B}$ is the bulk term. 
Since the one-point function of the perturbing operator does not vanish in the non-unitary CFT,
the first non-trivial coefficient in \eqref{eq:hatF} is the term with $k=1$.
In our case, this  is given by 
(see \cite{Zamolodchikov:1989cf,Klassen:1990dx} for example) 
\begin{align}
\hat{f}^{(1)}=\frac{\pi}{6} \hat{\kappa} \hat{C}^{(1)}, 
\label{eq:f1}
\end{align}
where
\begin{align}\label{C1}
\hat{C}^{(1)}=-12 (2\pi)^{\frac{n-8}{n}} C_{\hat{\Phi}_0 \hat{\Phi} \hat{\Phi}_0}, 
\end{align}
with  $\hat{\Phi}_0=\Phi_{\frac{n-3}{2},\frac{n-1}{2}}$
being the vacuum operator. 
$C_{\hat{\Phi}_0 \hat{\Phi} \hat{\Phi}_0}$ is the structure constant, 
which is given by \cite{Dotsenko:1985hi}
\begin{align}
(C_{\hat{\Phi}_0\hat{\Phi} \hat{\Phi}_0})^2
= \gamma\( -\frac{n-4}{n}\) \gamma\(-\frac{n-6}{n}\) \gamma \( \frac{4}{n}\)
\gamma^2 \( \frac{n-3}{n}\) \gamma^3 \( \frac{n-2}{n} \) \gamma^2 \( \frac{n-1}{n}\).
\end{align}
Substituting this into (\ref{C1}), we find
\begin{align}
\hat{f}^{(1)}=
4\pi \left[ \frac{1}{4\sqrt{\pi}} 
\frac{\Gamma(\frac{n}{4})}{\Gamma(\frac{n}{4}-\frac{1}{2})} \right]^{\frac{8}{n}}
\gamma\(\frac{4}{n}-1\)\gamma\( 1-\frac{3}{n}\)
\gamma^2\( 1-\frac{2}{n}\) \gamma\(1-\frac{1}{n}\).
\label{eq:free-T}
\end{align}

Next, taking into account the $\bbZ_{2}$-symmetry remarked above and
comparing the high-temperature expansions order by order, we obtain
\begin{align}
f_n^{(2k)}=2\hat{f}^{(k)}, \qquad f_n^{(2k+1)}=0 \qquad (k=1,2,\dots).
\label{eq:free-rel}
\end{align}
From  \eqref{eq:G-T}, we also find
\begin{align}
f_n^{(2)}=\frac{\pi}{6}C_n^{(2)}(\kappa_n^{\rm RSOS})^2 \cdot4\(1+\frac{F_{1,n-3}}{F_{11}}\)^2 .
\label{eq:free-T2}
\end{align}
Combining \eqref{eq:free-T}, \eqref{eq:free-rel} and \eqref{eq:free-T2}, 
we can fix the ratio $F_{1,n-3}/F_{11}$
 in $\kappa_{n} G$ as, 
\begin{align}
1+\frac{F_{1,n-3}}{F_{11}}=\frac{n-2}{n-3} \left[
\frac{\Gamma(\frac{n}{4})\Gamma(\frac{n-1}{2})}
{\Gamma(\frac{n}{4}-\frac{1}{2})\Gamma(\frac{n}{2})} \right]^{\frac{4}{n}} .
\end{align}
For $n=5$, this reproduces the result for the decagon considered in \cite{Hatsuda:2011ke}.
Once this ratio is fixed, we can obtain $\kappa_{n}G$ for general $M_{1}$ and $M_{n-3}$
with $M_{2} = \cdots =M_{n-4} =0$.

\setcounter{equation}{0}
\section{Comparison with two-loop results}

In this section, we compare the remainder function at strong coupling
in the previous sections with the two-loop results in 
\cite{Heslop:2010kq,Gaiotto:2010fk}.
By  numerically studying the remainder function
for $n=4$, it was noticed in \cite{Brandhuber:2009da} that appropriately
shifted and rescaled  remainder functions at strong coupling and at two loops
are close to each other. 
Such similarity was  observed also analytically 
in \cite{Hatsuda:2011ke} for $n=4$ and $n=5$.
Whether the similarity continues to hold for general $n$
would be a curious question, which should provide 
useful insights into the structure of the amplitudes. 
We thus discuss the case of the multi-point amplitudes below.

\subsection{Two-loop remainder function}

The analytic expression of the $2n$-point amplitudes 
has been given for the external momenta lying in a (1+1)-dimensional subspace 
of Minkowski space-time  \cite{Heslop:2010kq}, which correspond to the case of 
the minimal surfaces in AdS${}_{3}$. 
To write down the formula, we introduce the cross-ratios,
\eqb
  v_{ij}  \equiv \frac{x_{ij+1}^{2} x_{i+1j}^{2}}{x_{ij}^{2} x_{i+1 j+1}^{2}}  \period
\eqe
Denoting the cusp coordinates of the $2n$-gon as 
$x_{2i} = (x_{i}^{-},x_{i}^{+}) $, $ x_{2i+1} = (x_{i}^{-},x_{i+1}^{+})$,
$v_{ij}$ are reduced to $v_{ij} =1$ when $i-j$ is odd, and to
\eqb\label{uv}
  v_{2i+1,2j+1}  = u^{-}_{ij} \comma \qquad
  v_{2i,2j}=  u^{+}_{ij} \comma
\eqe
when $i-j$ is even, 
where $ 2 \leq | i-j | \ (\mbox{mod} \ n)  \leq n-2 $ and 
\eqb
  u_{ij}^{\pm}  
  \equiv \frac{x^{\pm}_{ij+1} x^{\pm}_{i+1j}}{x^{\pm}_{ij} x^{\pm}_{i+1 j+1}}  \period
\eqe
The remainder function then reads 
\eqb\label{2loopR}
 R_{2n}^{\rm 2\mbox{-}loop} = -\frac{1}{2} \sum_{\cal S}
  \log(v_{i_{1}i_{5}})\log(v_{i_{2}i_{6}})
    \log(v_{i_{3}i_{7}}) \log(v_{i_{4}i_{8}}) -\frac{\pi^{4}}{36}(n-2)\period
\eqe
The sum runs over 
\eqb
 {\cal S} = \Bigl\{ i_{1}, ..., i_{8} \bigm\vert  1\leq i_{1} < \cdots <
  i_{8} \leq 2n , \  \ i_{k} - i_{k-1} : {\rm odd}  \Bigr\} \period
\eqe
As on the strong-coupling side,  
the above formula of the two-loop remainder function preserves the $\bbZ_{2n}$-symmetry
(\ref{cyclic}) or (\ref{Z2n}).

In order to  compute the remainder function from the  Y-/T-functions,
we need to express $u^{\pm}_{ij}$ by $Y_{s}$/$T_{s}$. First, 
form (\ref{eq:yfunc1}) and (\ref{CR}) one finds that 
\eqb\label{uY}
 u^{+}_{k,-k-2} 
 =\frac{ Y_{2k+1}^{ [-1] } }{ 1+Y_{2k+1}^{ [-1 ] } }   
\comma \qquad 
 u^{+}_{k,-k-1} 
 =\frac{Y_{2k}^{{ [ 0 ]}} }{ 1+Y_{2k}^{{ [ 0 ]} } }
  \period
\eqe
Furthermore, the general $u^{+}_{ij}$ are obtained with the help of the
$\bbZ_{n}$-transformation (\ref{Zn}) induced by $Y_{s}^{[k]} \to Y_{s}^{[k+2]}$.
Namely, 
$u_{ij}^{+} 
= Y_{2k+1}^{ [ 2(i-k)-1 ] }/(1+ Y_{2k+1}^{  [ 2(i-k)-1 ]  })$
for $ j-i  = n-2k-2$, and 
$u_{ij}^{+} 
= Y_{2k}^{  [2(i-k)  ] }/(1+ Y_{2k}^{  [2(i-k)  ]  })$
for $  j-i  = n-2k-1$.
$u_{ij}^{-}$ are also obtained from $u_{ij}^{+}$ by the shifts $Y_{s}^{ [  k ]  } 
\to Y_{s}^{  [ k+1 ]  } $.
Eliminating $k$, 
we then arrive at the formulas,
\eqb\label{uYT}
  u_{ij}^{+} \Eqn{=} \frac{Y_{|i-j|-1}^{[i+j+1]}}{1+Y_{|i-j| -1}^{[i+j+1]}}
    = \frac{T_{|i-j| }^{[i+j+1]} T_{|i-j| -2}^{[i+j+1]} }{
      T_{|i-j| -1}^{[i+j+2]} T_{|i-j| -1}^{[i+j]}} \comma \nn \\
  u_{ij}^{-} \Eqn{=} \frac{Y_{|i-j|-1}^{[i+j+2]}}{1+Y_{|i-j| -1}^{[i+j+2]}}
    = \frac{T_{|i-j| }^{[i+j+2]} T_{|i-j| -2}^{[i+j+2]} }{
      T_{|i-j| -1}^{[i+j+3]} T_{|i-j| -1}^{[i+j+1]}}    
      \comma
\eqe
where $2 \leq | i - j | \leq n-2 $.
In the above, 
we have  used the symmetry $u_{ij}^{\pm} = u_{ji}^{\pm}$, 
the half-periodicity (\ref{Yperiod}) and the relations among $Y_{s}$
and $T_{s}$ in (\ref{T-system}), (\ref{YsTs}).

Using  the expansions of $T_{s}$ in (\ref{Texpand}), (\ref{Teven2}) and (\ref{That}), or
similar ones for $Y_{s}$, and 
the above expression of $u_{ij}^{\pm}$,  one can compute 
the expansion of the two-loop remainder function near the CFT limit.
As in the case at strong coupling, 
one then finds that the remainder function takes the form,
\eqb\label{Exp2loopR}
  R_{2n}^{\rm 2\mbox{-}loop} =   R_{2n}^{\rm 2\mbox{-}loop \, (0)}  
  +  l^{4(1-\Delta)}  R_{2n}^{\rm 2\mbox{-}loop \, (4)} + \calO( l^{6(1-\Delta)} ) \comma
\eqe
where $\Delta = (n-2)/n$.
By further using  the T-system (\ref{T-system}) or the Y-system (\ref{Y-system}),
$R_{2n}^{\rm 2\mbox{-}loop \, (4)}$ 
can be given again
by  $t_{1}^{(2,0)}\bar{t}_{1}^{(2,0)}$ or $y_{1}^{(2,0)}\bar{y}_{1}^{(2,0)}$,
where all the dependence of the mass parameters or the cross-ratios 
is included up to this order.

Since the remainder function in (\ref{2loopR}) contains an octuple sum,
the number of the terms in the sum rapidly increases.
We have computed the expansions up to $n=40$ 
for both odd and even $n$.  
In Table 1, we list  the numerical values 
of $R_{2n}^{\rm 2\mbox{-}loop \, (0)}$ 
and $R_{2n}^{\rm 2\mbox{-}loop \, (4)}/t_{1}^{(2,0)} \bar{t}_{1}^{(2,0)}$.

\begin{table}[t]
 \centering 
\begin{tabular}{c | | l | l | l || l | l | l || l }
& \multicolumn{3}{c||}{strong coupling} & \multicolumn{3}{c||}{2 loops} &  \ \, ratio
\\
\hline
$2n$ 
& $ \ \  R_{2n}^{(0)} $ & $ \quad \  r_{2n}^{ (4)}$  
& $  \quad \, \bar{r}_{2n}^{ (4)} $
& $ \ \ \, R_{2n}^{(0)}$ & $ \quad r_{2n}^{ (4)}$  
& $ \quad \, \bar{r}_{2n}^{ (4)} $
& $ \ \displaystyle \frac{ \bar{R}^{\rm strong}_{2n} }{  \bar{R}^{\rm 2\mbox{-}loop}_{2n} }$
\\
\hline\hline
$8$ &
\  $3.687$  \ &  $-0.003533$  & $-0.1638$  & 
\ $-5.527$ \  & $0.01843$ & $-0.1597$ &
\ $1.026$ 
\\
\hline
$10$ & 
\ $5.547$ \ & $-0.002183$ &  $-0.04419$ &
\ $-8.386$ \  & $0.01204$ &$-0.04490$& 
\ $0.9841$\\
\hline
$12$ &
\ $7.410$ \ &$-0.006482$ & $-0.08111$&
\ $-11.26$ \  & $0.03692$ &$-0.08441$& 
\ $0.9609$\\
\hline
$14$  & 
\ $9.275$ \ & $-0.01259$ & $-0.1126$ &
\ $-14.14$ \  & $0.07320$ &$-0.1190$& 
\ $0.9463$\\
\hline
$16$  & 
\ $11.14$ \  &$-0.02078$ & $-0.1437$  &
\ $-17.03$ \  & $0.1226$ &$-0.1535$& 
\ $0.9366$\\
\hline
$18$ & 
\ $13.01$ \ &$-0.03137$ & $-0.1764$ &
\ $-19.93$ \ & $0.1869$ &$-0.1897$&
\ $0.9297$\\
\hline
$20$  &
\ $14.87$ \ &$-0.04466$ & $-0.2112$&
\ $-22.82$ \  & $0.2682$ &$-0.2284$& 
\ $0.9247$\\
\hline
$22$  & 
\ $16.74$ \ & $-0.06098$&$-0.2486$&
\ $-25.72$ \  & $0.3683$ &$-0.2699$& 
\ $0.9209$\\
\hline
$24$  & 
\ $18.61$ \ & $-0.08063$ & $-0.2886$ &
\ $-28.61$ \  &$0.4892$ &$-0.3143$&
\ $0.9180$ \\
\hline
$30$ & 
\ $24.21$ \  & $-0.1626$ & $-0.4253$ &
\ $-37.31$ \  & $0.9955$ &$-0.4661$& 
\ $0.9124$\\
\hline
$50$ &
\  $42.88$ \ & $-0.7781$ & $-1.067$ &
\   $-66.32$ \ & $4.817$   & $-1.177$ & 
\ $0.9062$ \\
\hline
$80$ &
\ $70.89$ \ & $-3.220$ & $-2.571$ &
\ $-109.9$ \ & $20.02$ & $-2.843$& 
\ $0.9044$ \\
\hline
$200$ &
\ $182.9$ \  & $-50.60$ & $-15.10$ &
 \quad \ --  & \quad \ -- & \quad \ \ -- & 
\quad \ \  -- \\
\hline
$500$ &
\ $463.1$ \ & $-791.2$ & $-92.00$ &
\quad \ -- & \quad \ -- & \quad \ \ -- & 
\quad \ \  --  \\
\hline
$1000$ &
\ $930.0$ \ & $-6331$ & $-364.8$ &
\quad \ -- & \quad \ -- &\quad \ \ -- & 
\quad \ \  --  \\
\hline
\end{tabular}
\caption{
Expansion coefficients and ratios of the remainder functions at strong coupling 
and at two loops.
In the table, $r_{2n}^{(4)} \equiv R_{2n}^{(4)}/t_{s}^{(2,0)} \bar{t}_{s}^{(2,0)}$,
$\bar{r}_{2n}^{(4)} \equiv\bar{R}_{2n}^{(4)}/t_{s}^{(2,0)} \bar{t}_{s}^{(2,0)}$ for 
$2n \neq 8$, 
and 
 $r_{2n}^{(4)} \equiv R_{2n}^{(4)}$, 
$\bar{r}_{2n}^{(4)} \equiv \bar{R}_{2n}^{(4)}$ 
for $2n=8$. 
For large $n$, one finds that 
$R_{2n}^{\rm strong \, (0)} \approx 1.868 n$, 
$r_{2n}^{\rm strong \, (4)} \approx -5.065 \times 10^{-5} n^{3}$ and 
$\bar{r}_{2n}^{\rm strong \, (4)} \approx -1.447 \times 10^{-3} n^{2}$.
The values of $R_{2n}^{\rm 2\mbox{-}loop \, (0)}$ for $2n \leq 30 $ are
found in \cite{Brandhuber:2009da,Heslop:2010kq}, whereas the results 
 for $n=$ 4 and 5 are read off from \cite{Hatsuda:2011ke}. 
  (The case of $2n=8$ is special in that $t_{1}^{(2,0)}$ receives contributions
due to  the non-trivial $T_{n-2}$.)
}
\label{Table1}
\end{table}

\subsection{Rescaled remainder function}

Now, let us consider  the rescaled remainder function.
For the $2n$-point amplitudes, it  is defined by
\eqb
  \bar{R}_{2n} = \frac{R_{2n} -R_{2n,{\rm reg}}}{R_{2n,{\rm reg}} 
  - (n-2)R_{6,{\rm reg}}}
  \comma
\eqe
where $R_{2k,\rm{reg}}$ are the remainder functions  in the CFT limit corresponding to 
the regular $2k$-gons. For the hexagon, they are
\eqb
  R_{6,{\rm reg}}^{\rm strong} = \frac{7 \pi}{12} \comma \qquad 
  R_{6,{\rm reg}}^{\rm 2\mbox{-}loop} = - \frac{\pi^{4}}{36}  \comma
\eqe
at strong coupling and at two loops, respectively.
$\bar{R}_{2n}$ is calibrated so that it 
vanishes in the CFT limit and approaches $-1$ in the 
large $l$  limit when all the mass parameters are non-zero.
More generally,  if $k$ among $n-3$ mass parameters are zero,   
$R_{2n} \to (n-3-k) R_{6,{\rm reg}} + R_{2(k+3),{\rm reg}}$ for large $l$.
This is  understood by tracing the location 
of the poles of the polynomial $p(z)$ appearing in (\ref{eq:area1}),
and can be checked numerically.
In this case,  $R_{2n}$ approaches a constant different from $-1$.
In addition, when the mass parameters have a hierarchical structure, e.g., 
$m_{1}, m_{2} \gg m_{3} \gg \cdots$, the remainder function shows a plateau
at each scale where $m_{s}$ much smaller than that scale are 
regarded as effectively vanishing. The corresponding behavior
of  the Y-/T-functions  
have been studied in 
\cite{CastroAlvaredo:1999em,CastroAlvaredo:2000nr,Dorey:2004qc}.

One can then compute $\bar{R}_{2n}$ 
from the results at strong coupling in section 4 and 5, and those  at 
two loops in the previous subsection. 
They are  expanded as in the unrescaled case.
Up to $\calO(l^{4(1-\Delta)})$, they are  proportional to $t_{1}^{(2,0)}
\bar{t}_{1}^{(2,0)}$,
and the ratio $ { \bar{R}_{2n}^{\rm strong} }/{  \bar{R}_{2n}^{\rm 2\mbox{-}loop} }$
becomes a numerical number. In Table 1, we list the 
numerical values  for $\bar{R}_{2n}^{\rm strong}$ and 
$\bar{R}_{2n}^{\rm 2\mbox{-}loop}$ at $\calO(l^{4(1-\Delta)})$
divided by $t_{1}^{(2,0)} \bar{t}_{1}^{(2,0)} $, 
and their ratios up to $2n=80$.
We find that the remainder functions at strong coupling and at two loops 
continue to be close to each other for higher point amplitudes.
Furthermore, numerical plots for $2n= 12$ and $14$  show that the rescaled remainder 
functions are close at  any scale (Fig.~4), as observed for $2n=8$ and $10$ 
\cite{Brandhuber:2009da,Hatsuda:2011ke} .
We expect that this holds also for general $2n$-point amplitudes.

\begin{figure}[tbp]
\begin{center}
\begin{tabular}{cc}
\hspace{-3mm}
\resizebox{70mm}{!}{\includegraphics{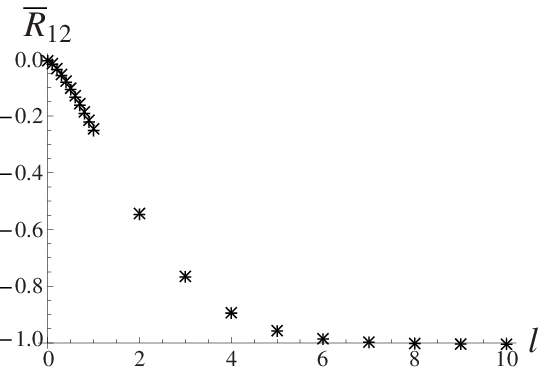}}
\hspace{3mm}
&
\resizebox{70mm}{!}{\includegraphics{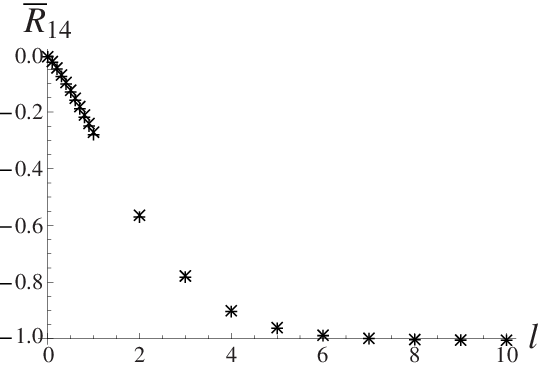}}
\hspace{-5mm}
\end{tabular}

\end{center}
\caption{Plots of the rescaled remainder functions at strong coupling $(\times)$ 
and at two loops
$(+)$
for 12-point amplitudes (left) and for 14-point amplitudes (right).
The functions are evaluated for $m_{s} = l e^{\frac{\pi i }{20}s} $.
At $l=2$, $\bar{R}_{12}^{\rm strong} =-0.538$, $\bar{R}_{12}^{\rm2\mbox{-}loop} =-0.542$,
whereas $\bar{R}_{14}^{\rm strong} =-0.559$, $\bar{R}_{14}^{\rm2\mbox{-}loop} =-0.565$.
}

\label{fig1}
\end{figure}

\subsection{Large $n$ limit}

Table 1 suggests that the ratio approaches a constant for  large $n$.
Let us consider this large $n$ behavior in more detail.\footnote{
In the large $n$ limit, the amplitudes are approximated by 
smooth Wilson loops after subtracting divergent terms
\cite{Alday:2007he,Alday:2009yn}.
}
In the following, we fix 
$\kappa_{n} G$ corresponding to the coupling in the CFT perturbation. 
 From (\ref{eq:ts20}), this implies
$t_{1}^{(2,0)} \sim \kappa_{n} G/n^{2}$ for large $n$.

On the strong coupling side,  the large $n$ behavior
of the remainder function is extracted from our formulas (\ref{R2nodd1})-(\ref{R2nodd3})
and (\ref{eq:R2neven1})-(\ref{eq:R2-even}).
At $\calO(l^{0})$, the summand $\log^{2}(t_{s}^{(0,0)}/t_{s-1}^{(0,0)})$ 
in $\Delta A_{\rm BDS}$
scales as $\frac{1}{n^{2}} h_{1}(x)$,
where $x=s/n$ and $h_{1}(x)$ is a certain function. 
Taking into account the fact that $h_{1}(1/n)$ grows as $(n \log 2)^{2}$,
one finds that  $\Delta A_{\rm BDS}$ at this order scales as 
$a_{1} n+ a_{0} + \cdots $. The other term at this order from 
$A_{\rm free}$, i.e., $c_{n}$,  has the same scaling. 
At $\calO(l^{\frac{8}{n}})$,  the terms in $A_{n,s}$
or $\hat{A}_{n,s}$  without 
$t_{s}^{(0,4)}$ scale as $\frac{1}{n^{2}} h_{2}(x)$, where $h_{2}(x)$ is
a certain smooth function. In addition, by numerically solving the 
equations for $t_{s}^{(0,4)}$ in terms of $t_{s}^{(2,0)}$,  one finds that 
the terms with $t_{s}^{(0,4)}$  also have the same scaling in $n$. 
Adding these terms,  $\Delta A_{\rm BDS}$ at this order  scales as 
$  \frac{b_{-1}}{n} +  \frac{b_{-2}}{n^{2}} + \cdots $.
The other term $f_{n}^{(2)}$ from $A_{\rm free}$ also has the same scaling.

Thus, the remainder function scales as $ (a'_{1} n+ a'_{0} + \cdots) 
+  l^{\frac{8}{n}}( \frac{b'_{-1}}{n} +  \frac{b'_{-2}}{n^{2}} + \cdots)$.
The linear behavior  at $\calO(l^{0})$ has been observed in \cite{Alday:2009yn}.
Indeed, by  a fit of the data for  $n = $   100 to 500 which 
includes terms up to  $\calO(n^{-3})$ at $\calO(l^{0})$ and up to  $\calO(n^{-5})$
at $\calO(l^{\frac{8}{n}})$, we find the
large $n$ behavior at strong coupling,
\eqb\label{RstrongScale}
 R^{\rm strong}_{2n} \Eqn{\approx} 
  1.868 n  \Bigl( 1- \frac{2.043}{n} + \frac{0.06340}{n^{2}} 
  + \frac{1.303 \times 10^{-7}}{n^{3}} 
 + \frac{0.08598}{n^{4}} \Bigr) \\
  && \quad 
  - \ l^{\frac{8}{n}} (\kappa_{n} G)^{2}  \cdot \frac{2.337}{n} 
  \Bigr( 1- \frac{6.703}{n} + \frac{24.90}{n^{2}} - \frac{62.32}{n^{3}} 
+ \frac{113.1}{n^{4}} \Bigr)  \period \nn
\eqe
The accuracy of the fit is of $\calO(10^{-13})$ at $\calO(l^{0})$
and of $\calO(10^{-11})$ at $\calO(l^{\frac{8}{n}})$.

On the two-loop side, we do not have a closed expression of the expansion of 
the remainder function. 
However, one can expect the same scaling as at strong coupling.
Indeed, the linear behavior at  $\calO(l^{0})$
has been observed   from numerical data up to $n=15$ \cite{Brandhuber:2009da}.
Furthermore, 
by performing a fit of the data for  $n =25 $ to 40
which includes terms up to  $\calO(n^{-3})$ at $\calO(l^{0})$ and up to  $\calO(n^{-5})$
at $\calO(l^{\frac{8}{n}})$, we find 
the large $n$ behavior at two loops,
\eqb\label{R2loopScale}
 R_{2n}^{\rm 2\mbox{-}loop} \Eqn{\approx} 
 -2.903 n \Bigl( 1- \frac{2.166}{n} + \frac{0.2544}{n^{2}} - \frac{0.001335}{n^{3}} 
 + \frac{0.4473}{n^{4}} \Bigr) \\
 && \quad 
 + \  l^{\frac{8}{n}}(\kappa_{n} G)^{2} \cdot \frac{14.57}{n} 
 \Bigl( 1- \frac{6.707}{n} + \frac{19.99}{n^{2}} - \frac{16.19}{n^{3}} 
 - \frac{66.06}{n^{4}} \Bigr) \period \nn
\eqe
The accuracy of the fit is of $\calO(10^{-8})$ at $\calO(l^{0})$ and 
of $\calO(10^{-10})$ at
$\calO(l^{\frac{8}{n}})$.
The behavior  at $\calO(l^{0})$ is consistent with the result 
in \cite{Brandhuber:2009da}.

In the fits,  the coefficient at  $\calO(n^{-2})$  in $R^{\rm strong}_{2n}$ is
of $\calO(10^{-7})$, which suggests that it is vanishing. In addition, the  term of 
$\calO(n^{0})$ 
in  $ R_{2n}^{\rm 2\mbox{-}loop} $ is 6.288 and close to $2 \pi$, as noted in 
\cite{Brandhuber:2009da}.
We also observe that at each order in $l$ the expansions 
in the parentheses in  (\ref{RstrongScale}) and (\ref{R2loopScale})
are similar to each other. 
It is expected that  they  become 
closer  with data for larger $n$ at two loops.

From the results  (\ref{RstrongScale}) and (\ref{R2loopScale}), 
we also find the ratio of the rescaled remainder functions for large $n$, 
\eqb
   \frac{ \bar{R}_{2n \gg 1}^{\rm strong} }{  \bar{R}_{2n \gg 1}^{\rm 2\mbox{-}loop} } 
   \approx 
   0.9049 - \frac{0.1178}{n} + \cdots  + \calO(l^{\frac{4}{n}}) \period
\eqe
Reflecting the similarity of the large $n$ expansion noted above, 
the ratio is close to 1, and 
the leading term is consistent with the expected value from Table 1.  
We note that a similar closeness has been observed between  
minimal surfaces in AdS and the amplitudes/Wilson loops at weak coupling
\cite{Galakhov:2008ax}.

\setcounter{equation}{0}
\section{Conclusions}

In this paper, we have studied the gluon scattering amplitudes of ${\cal N} = 4$ super
Yang-Mills theory at strong coupling by using the associated  Y-/T-system, 
focusing on the case where external momenta lie in a two-dimensional subspace 
$\bbR^{1,1}$.
In particular, by continuing the work  \cite{Hatsuda:2011ke}, 
we have considered  the analytic expansion of the $2n$-point  amplitudes
around  the momentum configurations  corresponding to the regular polygonal minimal
surfaces, or the high-temperature limit of the TBA system.

We  found that the cross-ratios $c_{i,j}^{\pm}, \hatc_{i,j}^{\pm}$, which 
appear in the remainder function, are concisely expressed in terms of  the T-function. 
This led to the simple expressions of  $\Delta A_{\rm BDS}$  (\ref{DelABDSTsodd}) and
(\ref{DelABDS}).
From  these expressions, 
we derived  the formulas (\ref{R2nodd1})-(\ref{R2nodd3})
and (\ref{eq:R2neven1})-(\ref{eq:R2-even}) for the leading-order expansion of 
the $2n$-point remainder function.  
The Y-/T-system   enabled us to encode 
its  momentum/mass-parameter dependence 
into only one function, e.g., $t_{1}^{(2,0)}$ in (\ref{eq:ts20}).
As shown in  \cite{Hatsuda:2011ke}, this function is computed  
by boundary CFT perturbation based on the relation between
the $g$-function (boundary entropy) and the T-function 
\cite{Bazhanov:1994ft,Dorey:1999cj}.
In addition to the result for $2n=10$ and those for general $n$ corresponding 
to the RSOS  scattering theory \cite{Hatsuda:2011ke},
we explicitly computed this function in the case
where  the TBA systems reduce to those
associated with the unitary and non-unitary diagonal coset CFTs.

We also compared our results at strong coupling with those at two loops 
\cite{Heslop:2010kq,Gaiotto:2010fk}.
As in the case of $2n=8,10$ \cite{Brandhuber:2009da,Hatsuda:2011ke},
the appropriately shifted and rescaled remainder functions
\cite{Brandhuber:2009da} 
continue to be close to each other for general
$n$.  Their ratio  at the leading order  tends to be a constant for large $n$.
Moreover, the original remainder functions at the leading order have similar $1/n$ expansions.

The observed closeness
suggests that the remainder function at general coupling 
is  constrained by some mechanism which is yet to be understood.
This would be an interesting issue for clarifying the full structure of the
amplitudes. 

It would also be interesting to extend our  analysis  to various directions. 
One is to find out the full mass-parameter dependence as in the case of 
$2n=10$ \cite{Hatsuda:2011ke}. Another is to derive the expansion
in the case corresponding to the minimal surfaces in AdS${}_{4}$  and AdS$_5$.
For these purposes, 
one needs to better understand multi-parameter integrable 
deformations of the CFTs 
associated with the relevant homogeneous sine-Gordon models.
In the general case of  AdS${}_{5}$,  the underlying integrable model and the CFT are 
not identified yet, in spite that its Y-system has been known\cite{Alday:2010vh}. 
This  would be an important future problem. 
One may also consider computation of higher order terms
in the expansion by extending  the boundary CFT perturbation in 
\cite{Dorey:1999cj,Dorey:2005ak} or by developing a formalism 
along the line of  \cite{Bazhanov:1994ft,BLZ,Fioravanti:2003kx}.

\vspace{3ex}

\begin{center}
{\large\bf Acknowledgments}
\end{center}

We would like to thank  J. Balog, A. Hegedus, K.  Sakai, J. Suzuki and R. Tateo
for useful discussions, and G. Korchemsky and S. Rey for useful comments.
Y. S. would also like to thank KFKI Research Institute for Particle and Nuclear
Physics, where part of this work was done, for its warm hospitality.
The work of K. I. and Y.~S. is supported in part by Grant-in-Aid
for Scientific Research from the Japan Ministry of Education, Culture, 
Sports, Science and Technology.

\vspace{3ex}
\appendix

\setcounter{equation}{0}
\renewcommand{\theequation}{\Alph{section}.\arabic{equation}}

\setcounter{equation}{0}
\section{Cross-ratios and  T-functions for even $n$}

In this appendix, we briefly summarize a procedure to relate 
the cross-ratios $\hatc^{\pm}_{i,j}$  and the T-functions 
for even $n$. As in the case of odd $n$, the relation is 
well understood graphically.

Let us first consider the cross-ratios consisting of $x_{k}^{+}$.
For $\hatc^{+}_{i,j}$ with odd $i-j$, 
namely, for $c_{i,j}^{aux \, +}$ with odd $i-j$, 
$c_{1,2k}^{right \, +}$ and $c_{1, 2k}^{left \, +}$,
discussion is similar to that in section 3.1.  We then find that 
\eqb\label{cAodd}
   c_{k+l,-k-1+l}^{aux \, +} \Eqn{=} 
    \prod_{p=0}^{n/2-k-2} \bigl(Y_{2k+1+2p}^{ [ 2l -1 ]} \bigr)^{(-1)^{p}}
    =  T_{2k}^{ [ 2l -1] } T_{n-2}^{ [ n+2(k+l)-1 ] } 
    \comma
\eqe 
where  $ k=1, ..., n/2-2$; $k+1 \geq l \geq 2-k$, and
\eqb\label{cRL}
   c_{1,-2k-2}^{right \, +} \Eqn{=} 
   \prod_{l=0}^{k} \bigl(Y_{2k+1-2l}^{ [ -2k-1 ]} \bigr)^{(-1)^{l}}
   = T_{2k+2}^{ [ -2k -1 ]} \comma \nn \\
   c_{1,-2k}^{left \, +} \Eqn{=} 
   \prod_{l=0}^{n/2-k-2} \bigl(Y_{2k+1+2l}^{ [ 1-2k ]} \bigr)^{(-1)^{l}}
   = T_{2k}^{ [ 1-2k ]} T_{n-2}^{ [ n+1 ] }\comma
\eqe
where $k=0, ..., n/2-2$.

The cross-ratios $\hatc_{i,j}^{+}$ with even $i-j$, namely, $c_{i,j}^{aux \, +}$ with even $i-j$, 
$d_{1,2k+1}^{right \, +}$ and $d_{1, 2k+1}^{left \, +}$,
are a little more complicated than those for odd $i-j$:
the first vertex  is dropped off or the edges stemming from the first or  the $n$-th
vertex are crossed (see Fig.~1 (b)-(d)).
This traces back to the fact  that $Y_{1}$ 
for the $2(n+1)$-point amplitudes is factored out non-trivially
in the double soft limit to the $2n$-point amplitudes \cite{Maldacena:2010kp}.
To write down the relation in this case,  it is helpful to use ``fan-shaped'' cross-ratios,
\eqb
  [ -k+1, k+2p-1, k+2p, k+2p+1 ]^{+} 
   \Eqn{=} \frac{1+Y_{2(k+p)-1}^{ [ 1+2p] }}{ Y_{2(k+p)-2}^{ [ 2p ] } }
      = \frac{ T_{2(k+p)-1}^{ [ 2+2p ] } }{T_{2(k+p)-3}^{ [ 2p ] }}
      \comma \nn \\
      {}[ k+1, -k-2p+3, -k-2p+2, -k-2p+1 ]^{+}
  \Eqn{=} \frac{1+Y_{2(k+p)-1}^{ [ 3-2p ] } }{ Y_{2(k+p)-2 }^{ [ 4- 2p ] } }
      = \frac{ T_{2(k+p)-1}^{ [ 2-2p ] } }{T_{2(k+p)-3}^{ [ 4-2p ] } } \period 
      \nn \\ 
\eqe
Combining these with 
$ c_{k+1,-k+1}^{aux \, +} = \prod_{l=0}^{k-2} 
\bigl(Y_{2k-2-2l}^{[ 2 ]} \bigr)^{(-1)^{l}} = T_{2k-1}^{ [ 2] } \bigl( T_{1 }^{ [ 2 ] } \bigr)^{(-1)^{k}}
$,
which are obtained similarly to (\ref{cAodd}), we find that
\eqb\label{cAeven}
  c_{k+1+2l,-k+1}^{aux \, +} \Eqn{=} 
          c_{k+1,-k+1}^{aux \, +} \prod_{p=1}^{l}
          \frac{ T_{2(k+p)-1}^{ [ 2+2p ] } }{T_{2(k+p)-3}^{ [ 2p ] }}
       = T_{2(k+l)-1}^{ [ 2+2l ] }  \bigl( T_{1 }^{ [ 2 ] } \bigr)^{(-1)^{k}}  
       \comma \nn \\
   c_{k+1,-k+1-2l}^{aux \, +} \Eqn{=} 
          c_{k+1,-k+1}^{aux \, +} \prod_{p=1}^{l}
        \frac{ T_{2(k+p)-1}^{ [ 2-2p ] } }{T_{2(k+p)-3}^{ [ 4-2p ] } }
       = T_{2(k+l)-1}^{ [ 2- 2l ] }  \bigl( T_{1 }^{ [ 2 ] } \bigr)^{(-1)^{k}} 
       \comma
\eqe
where  $k=1, ..., n/2 -1$; $l=0, ..., n/2 -k-1$.
In Fig.~5, we show a graphical 
representation of $c^{aux \, +}_{7,9} = c_{7,-1}^{aux \, +}$ 
for $n=10$, which  corresponds to the case of $k=l=2$ in the first equation in (\ref{cAeven}). 

\begin{figure}[tb]
\begin{center}
\resizebox{50mm}{!}{\includegraphics{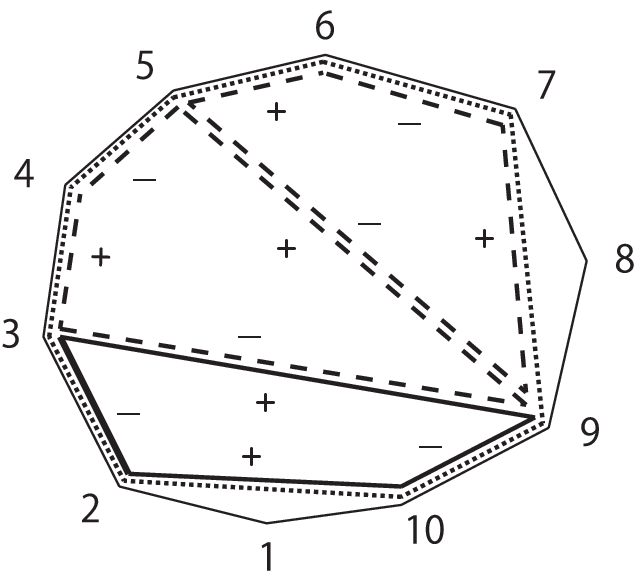}}
\end{center}
\caption{Decomposition of  $c^{aux \, +}_{7,9}$ for $n=10$.  
The $i$-th vertex stands for  $x_{i}^{+}$. The dotted line represents
$c^{aux \, +}_{7,9}$ and 
the bold line represents
the  tetragon corresponding to $c_{3,9}^{aux \, +}$. The dashed line represents 
the fan-shaped
tetragons corresponding to $T_{2(2+p)-1}^{[2+2p]}/T_{2(2+p)-3}^{[2p]}$ for $p=1,2$.
}
\label{fig:creven1}
\end{figure}

To find the expressions of $d_{1,2k+1}^{right \, +}$ and $d_{1,2k+1}^{left \, +}$,
we note that $d_{1,-2k+1}^{right \, +} = Y_{2k}^{[ -2k+2 ] }/c_{2,-2k}^{aux \, +}$
and $d_{1,2k+1}^{left \, +} = Y_{2k}^{[ 2k+2 ] }/c_{2(k+1),0}^{aux \, +}$.
It then follows that
\eqb\label{dRL}
  d_{1,-2k+1}^{right \, +} = T_{2k-1}^{[ 2-2k ]} T_{1}^{ [ 2] } \comma 
  \qquad 
  d_{1,2k+1}^{left \, +} = T_{2k-1}^{[ 2+2k ]} T_{1}^{ [ 2] }
  \comma
\eqe
where $k=1, ..., n/2 -1$. For $k=n/2 -1$, the intermediate relations to $Y_{2k}$
do not make sense. However, the above expressions hold also for $d_{1,3}^{right \, +}$
and $d_{1,-1}^{left \, +}$, which are obtained through
\eqb
  \frac{d_{1,2k+1}^{left \, +}}{d_{1,2k+1}^{right \, +}}
  =  \frac{c_{1,2k}^{right \, +}}{c_{1,2k}^{left \, +}} = c_{12}^{right \,+} 
  = T_{n-2}^{[ n -1] } \period
\eqe

The results in (\ref{cAodd})-(\ref{dRL})
cover all the non-trivial elements of $\hatc_{i,j}^{+}$. 
The corresponding expression  for $\hat{c}_{i,j}^{-}$ are obtained by the shift 
$T_{s}^{[ k]} \to T_{s}^{[ k +1 ]}$.
Furthermore, 
choosing the range of the indices as $1 \leq i,j \leq n+1$, 
they are summarized in the form given in the main text  (\ref{chatT}).
As in the course of the derivation, it is also possible to express
$\hatc_{i,j}^{\pm}$ by $Y_{s}$.

%
%
\def\thebibliography#1{\list
{[\arabic{enumi}]}{\settowidth\labelwidth{[#1]}\leftmargin\labelwidth
\advance\leftmargin\labelsep
\usecounter{enumi}}
\def\newblock{\hskip .11em plus .33em minus .07em}
\sloppy\clubpenalty4000\widowpenalty4000
\sfcode`\.=1000\relax}
\let\endthebibliography=\endlist
\vspace{3ex}
\begin{center}
{\large\bf References}
\end{center}
\par \vspace*{-2ex}

\end{document}